\documentclass{aa}
\usepackage[varg]{txfonts}
\usepackage[utf8]{inputenc}
\usepackage{amsmath}
\usepackage{amsfonts}
\usepackage{amssymb}
\usepackage{graphicx}
\usepackage{xcolor}
\usepackage{subfigure}
\graphicspath{ {./Plots/} }
\author{Rowan Batzofin \inst{\ref{inst1}}
\and Pierre Cristofari \inst{\ref{inst2}}
\and Kathrin Egberts \inst{\ref{inst1}}
\and Constantin Steppa \inst{\ref{inst1}}
\and Dominique M.-A. Meyer \inst{\ref{inst3}}}

\institute{Universit\"at Potsdam, Institut für Physik und Astronomie, Campus Golm, Haus 28, Karl-Liebknecht-Str. 24/25, 14476 Potsdam-Golm, Germany
 \label{inst1}
\and Observatoire de Paris, PSL Research University, 61 avenue de l’Observatoire, Paris, France \label{inst2}
\and Institute of Space Sciences (ICE, CSIC), Campus UAB, Carrer de Can Magrans s/n, 08193 Barcelona, Spain \label{inst3}} 

\usepackage{aas_macros}
\usepackage{url}
\usepackage{hyperref}
\usepackage{natbib}
\bibpunct{(}{)}{;}{a}{}{,}
\DeclareUnicodeCharacter{03B3}{$\gamma$}
\usepackage{xcolor}
\usepackage{siunitx}

\begin{document}

\title{The population of Galactic supernova remnants in the TeV range}

\abstract {Supernova remnants (SNRs) are likely to be significant sources of cosmic rays up to the knee of the local cosmic-ray (CR) spectrum. They produce gamma-rays in the very-high-energy (VHE) ($E>0.1$ TeV) range mainly via two mechanisms: hadronic interactions of accelerated protons with the interstellar medium and leptonic interactions of accelerated electrons with soft photons. Observations with current instruments have lead to the detection of about a dozen of SNRs emitting VHE gamma-rays and future instruments should significantly increase this number. Yet, the details of particle acceleration at SNRs, and of the mechanisms producing VHE gamma-rays at SNRs remain poorly understood.} 
{
We aim at studying the population of SNRs detected in the TeV range and its properties, and confront it to simulated samples, in order to address fundamental questions of particle acceleration at SNR shocks: 
 What is the spectrum of accelerated particles? What is the efficiency of particle acceleration? Is the gamma-ray emission dominated by hadronic or leptonic interactions?} 
{By means of Monte Carlo methods, we simulate the population of SNRs in the gamma--ray domain and confront our simulations to the catalogue of sources from the systematic survey of the Galactic plane performed by H.E.S.S. (HGPS).}
{We systematically explore the parameter space defined in our model, including e.g., the slope of accelerated particles $\alpha$, the electron-to-proton ratio $K_{\rm ep}$, and the  efficiency of particle acceleration $\xi $. In particular, we found possible sets of parameters for which $\gtrsim 90$\% of Monte Carlo realisations are found in agreements with the HGPS data. These parameters are typically found $ 4.2 \gtrsim \alpha  \gtrsim 4.1 $, $10^{-5} \lesssim K_{\rm ep} \lesssim 10^{-4.5}$, and $0.03 \lesssim \xi\lesssim 0.1 $ . We were able to strongly argue against some regions of the parameter space, such as e.g. $\alpha \lesssim 4.05$, $\alpha \gtrsim 4.35$, or $K_{\rm ep} \gtrsim 10^{-3}$. 
}
{Our model is so far able to explain the SNR population of the HGPS. 
Our approach, confronted to the results of future systematic surveys, such as with the Cherenkov Telescope Array,  will help remove 
degeneracy in the solutions, and to better understand 
particle acceleration at SNR shocks in the Galaxy.
}

\maketitle
\section{Introduction}
Cosmic rays (CRs) were discovered thanks to the pioneer work of Domenico Pacini and Victor Hess in the 1910s and have been extensively studied ever since  \citep{Hess:1912srp}. Remarkably, the most  fundamental question of CR physics remains unanswered: where are they produced? The standard paradigm for the origin of Galactic CRs is that supernova remnants (SNRs) accelerate the bulk of CRs, via the first-order Fermi mechanism called diffusive shock acceleration (DSA) \citep{2010ApJ...718...31P}. DSA can efficiently energise protons (and ions), as well as electrons. Subsequently, the accelerated protons and electrons can interact with the interstellar medium (ISM) to produce gamma-rays in the very-high-energy (VHE) domain, mainly through two mechanisms: 1) the production/decay of neutral pions, in the collision of accelerated protons with hadrons of the ISM (hadronic mechanism), and 2) the inverse Compton scattering of accelerated electrons on soft photons: cosmic microwave background (CMB), infrared, optical (leptonic mechanism).
Numerous SNR shocks have now been detected in the gamma-ray domain, with instruments such as Fermi-LAT~ \citep{Fermi2016} in the GeV range and in the TeV range with Cherenkov instruments such as VERITAS~ \citep{VERITAS_SNR}, MAGIC~ \citep{MAGIC_SNR} and H.E.S.S.~ \citep{HESS_SNRPop2018}.

While detailed studies of individual objects have been performed extensively in the past (e.g. RX J1713-3946  \citep{RX_J1713}, Cas A  \citep{Cas_A}, Vela Jr  \citep{Vela_Jr}, SN 1006  \citep{Pais_2020}), the study of the SNR population itself can unveil valuable information.

 In particular, H.E.S.S. (the High Energy Stereoscopic System), an array of imaging atmospheric Cherenkov telescopes located in the Khomas highlands in Namibia,  performed a systematic survey of the Galactic plane. The H.E.S.S. Galactic Plane Survey (HGPS) covers the Galactic plane from longitudes of $250^{\circ}$ to $65^{\circ}$ and latitudes of $-3^{\circ}$ to $3^{\circ}$. This systematic survey found 78 sources in the TeV domain, 8 were clearly detected SNRs and 47 sources were classified as unidentified.

In this work, we intend to simulate the population of TeV emitting SNRs and confront it to the population detected in the HGPS. We follow the framework presented in \citet{Cristofari_2013} (C13) and build on it with a refined modeling and a comparison with the HGPS SNR catalogue based on the method introduced in \citet{Steppa_2020}. 
 
 For this purpose we simulate populations of Galactic SNRs using a Monte Carlo method. We simulate the explosion times and positions of the supernovae and calculate the gamma-ray emission from the SNRs at the current time. We constrain the simulations with measurements from the HGPS by identifying the detectable sources in the simulations based on the longitude, latitude, and angular-extension dependent HGPS sensitivity and comparing it with the detected sample of SNRs in the HGPS. The inclusion of the complex dependencies of the HGPS sensitivity is crucial for drawing conclusions on the SNR population based on this data comparison. Due to the large number of unidentified sources (47) in the HGPS, the SNRs that are identified as such in the HGPS (8) are treated as a lower limit. We use as a stringent upper limit (28) the sum of firmly identified SNRs (8), firmly identified composite objects (8), and of unidentified sources associated (by position) with a known SNR (12). For a strict upper limit (63) we include all the sources in the stringent upper limit as well as the rest of the unidentified sources (35).

 With respect to C13, the present work especially includes several refined aspects on the modelling, such as: 1) inclusion of a connection between the ejecta masses and explosion energies for the SNRs; 2) refined description of the magnetic field amplification and corresponding maximum energy for protons and electrons; 3) inclusion of diffusive shock reacceleration at SNR shocks; 4) discussion of several prescriptions for the spatial distribution of SNRs in the Galaxy. 

\paragraph{}
The paper is structured as follows: in Sect. \ref{section:pop_model} we describe the population model: the physics of individual SNRs and the distributions used to place them in the Milky Way. In Sect.~\ref{section:simulating_populations} we discuss the methods for comparing our simulations to the HGPS, and present our results in Sect.~\ref{section:results}. We present our conclusions in Sect.~\ref{section:conclusion}.

\section{SNR population model} \label{section:pop_model}

Following~C13, in order to simulate the population of Galactic SNRs emitting in the TeV range, we need: a) a model for particle acceleration at SNRs and the subsequent gamma--ray emission (from hadronic and leptonic interactions of accelerated particles with the surrounding medium), which needs to be computationally inexpensive; b) a description of the spatial distribution of SNRs in the Galaxy; c) a description of the gas density distribution in which the SNR shocks are expanding. These three ingredients are discussed in the following subsections. 

\subsection{Modeling the gamma-ray emission of SNRs} \label{section:gamma-ray_calc}

\subsubsection{Particle acceleration at SNR shocks}
\label{section:CR_acc}
We rely on the usual assumption that a fraction $\xi_{\rm CR}$ of the SNR shock ram pressure $\rho u_{\rm sh}^2$ is converted into accelerated particles (loosely referred to as CRs in the following). 
To this end, we assume that the CR particle distribution at a time $t$ and at a position $r$ inside the shock follows a power law in momentum, as expected with DSA, such that $f_{\rm CR}(R_{\rm sh}, p, r) = A(r, t)\left(\frac{p}{m_p c}\right)^{-\alpha}$, where $p$ is the particle momentum and $\alpha$ is treated as a free parameter in the range $\alpha \geq 4$.  We can find the normalisation $A$ by requiring the cosmic-ray pressure to be equal to some fraction, $\xi_{CR}$, of the shock ram pressure at the shock location, $\rho u^2_{sh}$:
\begin{equation}
\frac{1}{3} \int_{p_{\text{min}}}^{p_{\text{max}}} \: \text{d}p \: 4\pi p^2 f_{\rm CR}(p) \: pv(p) = \xi_{\rm CR} \rho u^2_{sh} \,,
\label{equation:CR=ram}
\end{equation}
where 
\begin{equation}
A = \frac{3}{4\pi}\frac{\xi_{\rm CR} \rho u_{\rm sh}^2}{m_p^4 c^5 I(\alpha)} \, 
\label{equation:normalisation}
\end{equation} 
\noindent 
and
\begin{equation}
I(\alpha) = \int_{\frac{p_{\rm min}}{m_pc}}^{\frac{p_{\rm max}}{m_pc}}dx,  \frac{x^{4-\alpha}}{\left(1+x^2\right)^{1/2}} \, .
\end{equation} 
The spatial distribution of cosmic rays inside the SNR can be computed by solving the transport equation:
\begin{equation}
\frac{\partial f}{\partial t} + u \nabla f - \nabla D \nabla f - \frac{p}{3} \nabla u \frac{\partial f}{\partial p} = 0 \, ,
\label{equation:transport}
\end{equation}
where $D$ is the momentum-dependent diffusion coefficient for cosmic rays. Since particles with momenta $< p_{\rm max}$ are expected to be well confined within the SNR, the diffusion term $(\nabla D \nabla f)$ is much smaller than the advection term $(u \nabla f)$. Eq. \ref{equation:transport} can then be solved using the method of characteristics and using the boundary condition: $f(R_{\rm sh}, p, t) = f_0 (p, t)$.

\subsubsection{Particle reacceleration at SNR shocks}
\label{sec:reac}
To determine the contribution of gamma-ray emission from the reaccelerated particles we follow \citet{Cristofari_2019}. The contribution from the reaccelerated particles is given by:
\begin{equation}
    f_{0}^{\text{seed}}(p) = \alpha \int_{p_0}^p \frac{dp'}{p'}\left(\frac{p'}{p}\right)^{\alpha} f_{\infty}(p') \, ,
    \label{equation:reacc_spec}
\end{equation}
where $\alpha$ is the spectral index as before, $f_{\infty}(p)$ is the distribution function at upstream infinity of the seeds to be reaccelerated, in this case Galactic CRs, and $p_0$ is the minimum momentum of reaccelerated particles. The exact value of $p_0$ is not critical for Galactic CRs as discussed in \citet{Blasi_2017}, in this case we use $p_0 = 10^{-2}$ GeV. We assume that the spectrum of Galactic protons and electrons is the same as the local interstellar spectrum and use the parameterisations as done in \citet{Bisschoff_2019}, which describe the Voyager~I \citep{Cummings_2016} and PAMELA \citep{Adriani_2011} data well.

\subsubsection{Acceleration of electrons}

Just as with protons, DSA is also efficiently energising electrons. The main difference is that the electrons suffer synchrotron and inverse Compton losses, thus affecting the spectra of accelerated particles. At low energies where these losses can be neglected the spectra of protons and electrons have the same shape. We introduce a parameter, $K_{\rm ep}$, to describe the ratio of electrons to protons at these low energies. The maximum energy of electrons at the shock is found by equating the acceleration time at the shock to the synchrotron energy loss time. Following the approach of \citet{Vannoni_2009}, the maximum energy of electrons is:

\begin{equation}
E_{\rm max}^e \approx 7.3 \left(\frac{u_{\rm sh}}{1000 \text{ km/s}}\right) \left(\frac{B_{\rm down}}{100 \text{ }\mu \text{G}}\right)^{-1/2} \text{ TeV}. 
\label{equation:Emax}
\end{equation}

The electrons are accelerated at time scales much shorter than the synchrotron loss time. After being accelerated the electrons are advected downstream of the shock where they continue to lose energy through synchrotron radiation. The characteristic time for synchrotron radiation is:

\begin{equation}
\tau_{syn} \approx 1.8 \times 10^3 \left(\frac{E_e}{TeV}\right)^{-1} \left(\frac{B_{\rm down}}{100 \text{ } \mu \text{G}}\right)^{-2}\text{ yr}, 
\end{equation}
where $E_e$ is the electron energy. The energy loss time decreases with electron energy and there could be an energy ($E_{\rm break}^e$) beyond which the loss time is less than the age of the SNR. Above this energy the electron spectrum is shaped by radiative losses and steepens by one power in momentum  \citep{Morlino_2012}. The SNR is expanding so there are also adiabatic losses with a rate of $\tau_{\rm ad} = \frac{R_{\rm sh}}{u_{\rm sh}}$. Hence, 
$E_{\rm break}^e$ can be found by solving the following equation \citep{Finke_2012}: 
\begin{equation}
\tau_{\rm age}^{-1}=\tau_{\rm syn}^{-1}+\tau_{\rm ad}^{-1}\, .
\label{equation:E_break}
\end{equation}

\subsubsection{Maximum energy of accelerated particles}

The question of the maximum energy of accelerated particles at SNR shocks is fundamentally connected to the question of the amplification of magnetic fields at SNR shocks. For fast SNR shocks, the amplification of magnetic fields is expected to be determined by the growth of non-resonant streaming instabilities upstream the SNR shock~ \citep{Bell_2013,schure2013,Cristofari_review_2021}.
As the SNR shock slows down, resonant streaming instabilities become relevant \citep{schure2012}. We account for the magnetic field amplification as in \citet{Cristofari_2021}.  For the resonant streaming the upstream magnetic field is estimated as

\begin{equation}
    B_{\rm up}^2 \approx \frac{1}{M_A}\xi_{CR} v_{\rm sh}^2 \, , 
    \label{equation:B_resonant}
\end{equation}
where $M_A = v_{\rm sh}/v_{\rm A,0}$ is the Alfv\' enic Mach number.
\paragraph{}
In the non-resonant case the magnetic field stops growing when the energy density in the form of escaping particles equals that in the amplified field:
\begin{equation}
    \frac{\delta B^2}{4 \pi} \approx 3 \frac{v_{\rm sh}}{c} \frac{\xi_{\rm CR} \rho v^2_{\rm sh}}{(\alpha-3) I(\alpha)} \left( \frac{p_{\rm min}}{m_pc}\right)^{4-\alpha}. 
    \label{equation:B_Bell}
\end{equation}

\paragraph{}
Assuming the turbulent field upstream of the shock is roughly isotropic and the perpendicular components are compressed at the shock, the downstream magnetic field is given by: $B_{\rm down} \approx B_{\rm up} \sqrt{\frac{1+2r^2}{3}}$, where $r$ is the compression factor (set to 4 in this work). The amplified magnetic field for the SNR is given by: $\sqrt{B_{\rm resonant}^2 + B_{\rm non-resonant}^2+B_0^2}$, where $B_0$ is the magnetic field in the interstellar medium ($3 \mu G$ in this work)~\citep{ferriere2001}. 

\paragraph{}
Two different methods were used to estimate the maximum energy of the protons accelerated at the shock. The first method (which we will refer to as Bell) follows the work in \citet{Bell_2013} and \citet{Cristofari_2020} where the maximum energy of the protons is estimated using the following condition: $\int_0^t dt' \gamma_{\rm max}(t') \approx 5$, where $\gamma_{\rm max}$ is the maximum non-resonant hybrid growth rate, which leads to:  
\begin{equation}
    \left( \frac{p_{\rm max}}{m_p c}\right)^{\alpha-3} = \frac{e \: R_{\rm sh} \: B_{\rm up}^2}{20 \sqrt{\pi \rho} \: v_{\rm sh} m_p c^2 } \left(\frac{p_{\rm min}}{m_p c} \right)^{\alpha-4} \, .
    \label{equation:pmax_Bell}
\end{equation}
The second method (which we will refer to as Hillas) follows C13 where the maximum momentum of the accelerated protons, $p_{\text{max}}$, is determined by the following equation: 
\begin{equation}
l_d = \frac{D(p_{\text{max}})}{u_{\rm sh}} \approx \zeta R_{\rm sh}\, , 
\label{equation:pmax}
\end{equation}
\noindent where $D$ represents the momentum-dependent diffusion coefficient for the CRs upstream of the shock,  $l_d$ is the diffusion length of the shock and $\zeta$ is some fraction, we adopt $\zeta = 0.1$ here. The diffusion coefficient depends on the magnetic field strength and its structure. We assume that the the CR diffusion coefficient is of the Bohm type and $D = \frac{R_L c}{3}$, where $R_L = \frac{pc}{qB}$ is the particle Larmor radius, $q$ is the elementary charge, $c$ is the speed of light, and $B$ is the amplified magnetic field. 
\paragraph*{}
We use a diffusion coefficient for the cosmic rays at the shock \citep{Ptuskin_2012}: 
\begin{equation}
D = D_B \left(1+\frac{v^2_d}{u_{\rm sh}^2}\right)^{\rm g},
\label{equation:D}
\end{equation}
where $D_B$ is the Bohm diffusion 
coefficient, $v_d$ is the velocity that defines the importance of wave damping in limiting the field of amplification, and the parameter $g$ depends on the nature of the dominant damping mechanism. We use $g = 1.5$ and following \citet{Zirakashvili_2010} adopt the following expression for $v_d$:
\begin{equation}
    v_d = \left(\frac{\sigma^2 B_{up}^2}{8\pi \xi_B \varrho_{up}}\right)^{1/2},
\label{equation:v_d}
\end{equation}
where  $\varrho_{up}$ represents the gas density upstream of the shock and $\xi_B$ is the fraction of the shock ram pressure, $\varrho_{up} u_{sh}^2$, converted into magnetic field. \citet{Vokl_2005} found that $\xi_B \approx 3.5\%$ by looking at X-ray data form young SNRs and we use that value in this work.
Plugging all this information into Eq. \ref{equation:pmax} we can solve for the maximum energy of protons.

\subsubsection{Dynamics of the SNR shocks}
\label{section:shock_evolution}

In the description of particle acceleration and of the maximum energy of accelerated particles, the dynamics of the SNR shocks, radius ($R_{\rm sh}$) and speed ($u_{\rm sh}$), plays an important role.

We follow the approach of~\citet{Ptuskin_2003, Ptuskin_2005} where it is assumed that there is very efficient acceleration and the cosmic ray pressure inside the supernova governs its expansion. For thermonuclear SNRs the shock speed and radius in the ejecta-dominated phase is described by \citep{Chevalier_1982,Ptuskin_2005}:
\begin{equation}
R_{\rm sh} = 5.3 \left(\frac{\epsilon^2_{51}}{n_0\: M_{\rm ej,\odot}}\right)^{1/7} t_{\rm kyr}^{4/7} \text{ pc}
\label{equation:Ia_ejecta_radius}
\end{equation}
and
\begin{equation}
u_{\rm sh} = 3.0 \times 10^3 \left(\frac{\epsilon^2_{51}}{n_0\: M_{\rm ej,\odot}}\right)^{1/7} t_{\rm kyr}^{-3/7} \text{ km/s}\, , 
\label{equation:Ia_ejecta_speed}
\end{equation}
\noindent where $\epsilon_{51}$ is the explosion energy in $10^{51}$ erg, $n_0$ is the ambient gas density in cm$^{-3}$, $M_{\rm ej,\odot}$ is the mass of the ejecta in solar mass units and $t_{\rm kyr}$ is the time after the explosion in kyr.

\paragraph*{}
The shock speed and radius in the adiabatic phase can be calculated with the following equations
\citep{Truelove_1999, Ptuskin_2005}:
\begin{equation}
R_{\rm sh} = 4.3 \left(\frac{\epsilon_{51}}{n_0}\right)^{1/5} t_{\rm kyr}^{2/5} \left(1 - \frac{0.06\: M_{\rm ej, \odot}^{5/6}}{\epsilon_{51}^{1/2}\: n_0^{1/3}\: t_{\rm kyr}}\right)^{2/5} \text{ pc}
\label{equation:Ia_adiabatic_radius}
\end{equation}
and
\begin{equation}
u_{\rm sh} = 1.7 \times 10^3 \left(\frac{\epsilon_{51}}{n_0}\right)^{1/5} t_{\rm kyr}^{-3/5} \left(1 - \frac{0.06 \: M_{\rm ej, \odot}^{5/6}}{\epsilon_{51}^{1/2}\: n_0^{1/3}\: t_{\rm kyr}}\right)^{-3/5} \text{ km/s},
\label{equation:Ia_adiabatic_speed}
\end{equation}
respectively.
\noindent Eqs. \ref{equation:Ia_ejecta_radius} and \ref{equation:Ia_ejecta_speed} connect smoothly to Eqs. \ref{equation:Ia_adiabatic_radius} and \ref{equation:Ia_adiabatic_speed} at the transfer time $t_{\rm transfer} \approx 260\left(\frac{M_{\rm ej,\odot}}{1.4}\right)^{\frac{5}{6}} \epsilon_{51}^{-\frac{1}{2}} n_0^{-\frac{1}{3}}$ yr.

\paragraph*{}
Core-collapse supernova shocks propagate in the wind-blown bubble generated by the wind of the progenitor star \citep{das_aa_661_2022}. The wind-blown bubble is divided into two regions: a dense red-super-giant wind and a tenuous hot bubble inflated by the wind of the progenitor star. We assume the gas density distribution of the wind-blown region is spherically symmetric and that the stellar wind has a velocity $u_w = 10^6$ cm/s. The mass loss rate is of the order of $\dot{M} = 10^{-5} \dot{M}_{-5}$, where $\dot{M}_{-5}$ is the wind mass loss rate \citep{meyer_mnras_506_2021}. The density of the gas in the wind is $n_w = \frac{\dot{M}}{4 \pi m_a u_w r^2}$, where $u_w$ is the stellar wind velocity, $r$ is the radius of the wind and $m_a = \mu m_p$ is the mean interstellar atom mass ($\mu$ is assumed to be 1.4 here). The density of the hot bubble is  \citep{Weaver_1977}: $n_b = 0.01\left(L^6_{36}n_0^{19}t_{\rm Myr}^{-22}\right)^{1/35} \text{ cm}^{-3}$, where $L_{36}$ is the star wind power in units of $10^{36}$ erg $s^{-1}$ ($L_{36} \sim 1$) and $t_{\text{Myr}}$ is the wind lifetime in units of Myr ($t_{\text{Myr}} \sim 1$) \citep{Longair_1994}. The radius of the wind is set by equating the wind pressure to the thermal pressure of the hot bubble: $r_w = \sqrt{\frac{\dot{M} u_w}{4\pi k n_b T_b}}$, where $k$ is the Boltzmann constant and $T_b = 1.6 \times 10^6 \left(n_0^2 L^8_{36} t_{\text{Myr}}^{-6}\right)^{\frac{1}{35}}$~K  \citep{Castor_1975}. The radius of the hot bubble is: $r_b = 28 \left(\frac{L_{36}}{\mu n_0}\right)^{1/5}t_{\text{Myr}}^{3/5}$~pc. 

\paragraph*{}
For core collapse SNRs the ejecta-dominated phase is described by the following equations \citep{Chevalier_1982, Ptuskin_2005}:
\begin{equation}
R_{\rm sh} = 7.7 \left(\frac{\epsilon_{51}^{7/2}\: \frac{u_{w}}{10^6}}{\dot{M}_{-5}\: M_{\rm ej,\odot}^{5/2}}\right)^{1/8} t_{\rm kyr}^{7/8} \text{ pc}\, 
\label{equation:cc_ejecta_radius}
\end{equation}
and
\begin{equation}
u_{\rm sh} = 6.6 \times 10^3  \left(\frac{\epsilon_{51}^{7/2}\: \frac{u_{w}}{10^6}}{\dot{M}_{-5}\: M_{\rm ej,\odot}^{5/2}}\right)^{1/8} t_{\rm kyr}^{-1/8} \text{ km/s} \, .
\label{equation:cc_ejecta_speed}
\end{equation}

\noindent To describe the evolution of core-collapse SNRs in the adiabatic phase, we adopt the thin-shell approximation, implying that the gas swept up by the shock is concentrated in a thin layer behind the shock front. Using this we can describe the shock speed and the age of the SNR as functions of the shock radius using the following equations \citep{Ptuskin_2005}:

\begin{align}
\begin{split}
u_{\rm sh}(R_{\rm sh}) = \frac{\gamma_{\rm ad} + 1}{2} & \left[\frac{12(\gamma_{\rm ad}+1)\epsilon}{(\gamma_{\rm ad}-1)M^2(R_{\rm sh})R_{\rm sh}^{6(\gamma_{\rm ad}-1)/(\gamma_{\rm ad}+1)}} \right. \\
& \left. \times \int_0^{R_{\rm sh}}{dr\: r^{6 \left( \frac{\gamma_{\rm ad}-1}{\gamma_{\rm ad}+1} \right)-1}M(r)} \right]^{1/2} \, 
\end{split}
\label{equation:cc_adiabatic_speed}
\end{align}
and
\begin{equation}
t(R_{\rm sh}) = \int_0^{R_{\rm sh}}{\frac{dr}{u_{\rm sh}(r)}} \, ,
\label{equation:cc_adiabatic_time}
\end{equation}
\noindent where $\gamma_{\rm ad}$ is the gas adiabatic index and $M$ is the total gas mass (ejecta and swept-up) inside the shock: 
\begin{equation}
M(R) = M_{\text{ej}}\: +\: 4\pi \: \int_0^R dr\: r^2 \: \rho (r) \,,
\label{equation:mass}
\end{equation}
\noindent where $M_{\text{ej}}$ is the ejected mass and $\rho$ is the density of 
ambient gas.

When the mass of the swept up gas becomes larger than the mass of the ejecta, the SNR enters the adiabatic (Sedov-Taylor (ST)) phase (in this work we assume that this is when the mass of the swept up gas is twice the mass of the ejecta). Eqs. \ref{equation:cc_ejecta_radius} and \ref{equation:cc_ejecta_speed} are multiplied by normalisation constants such that at the transfer time they match the shock radius and shock speed calculated using Eqs. \ref{equation:cc_adiabatic_speed} and \ref{equation:cc_adiabatic_time}.

\paragraph*{}
This description of the SNR shock holds until the end of the ST phase, which is assumed to be when the acceleration of particle stops being efficient. The end of the ST phase can typically be estimated by equating the cooling time to the age of the SNR~ \citep{Blondin_1998}:
\begin{equation}
t_{\rm cool} \approx 10^3 \left(\frac{n_0}{1\: \text{cm}^{-3}}\right)^{-1} \left(\frac{v_{\rm sh}(t)}{10^8\: \text{cm}\:\text{s}^{-1}}\right)^3 \: \text{kyr} \,.
\label{equation:cooling_time}
\end{equation}

\subsubsection{Gamma rays}
The gamma--ray emission from accelerated (and reaccelerated) protons and electrons can be computed, making use of the {\sc Naima} Python package  \citep{naima}.
The hadronic emission is calculated using the pion decay model \citep{naima_PD}, the proton spectrum is described by a power law with exponential cut-off. The normalisation is calculated using Eq. \ref{equation:normalisation}, the cut-off energy is $p_{\rm{max}}$, calculated using either Eq. \ref{equation:pmax_Bell} or Eqs. \ref{equation:pmax}-\ref{equation:v_d} and the power law index is $\alpha -2$ since $\alpha$ is the power law in momentum. The leptonic emission is calculated using the inverse Compton model \citep{naima_IC} of the {\sc Naima} package, with only the CMB as a radiation field. The electron spectrum is described by a broken power law with exponential cut-off. This is calculated similarly to the protons except the normalisation is multiplied by the electron to proton ratio ($K_{\rm ep}$) and the break energy is calculated by solving Eq. \ref{equation:E_break}, the power law index after the break is $\alpha -1$ and the cut-off energy is calculated with Eq. \ref{equation:Emax}. The total simulated flux for an individual SNR is the sum of the hadronic emission and the leptonic emission from both the accelerated and reaccelerated protons and electrons.

\subsection{Distribution of Galactic SNRs}

\label{section:source_distribution}

\subsubsection{Distribution of the physical parameters}
The simulated SNRs are drawn relying on a Monte Carlo approach: the age of the SNRs is taken following a uniform distribution and we assume a constant supernova explosion rate of 3 per century. We account for SNRs from both thermonuclear (32\%) and core-collapse (68\%) supernovae (SNe) \citep{Smart_2009}. For SNRs from thermonuclear SNe the total explosion energy is $10^{51}$ erg and the mass of the ejecta ($\text{M}_{\text{ej}}$) is 1.4 $\text{M}_{\odot}$. For SNRs from core-collapse SNe the masses of the ejecta depend on the zero-age main-sequence mass of the star, the explosion energies depend on the mass of the ejecta. The progenitor's last mass loss rate is assumed to be that of a red supergiant star, i.e.  $\dot{\text{M}} =10^{-5}$ $\text{M}_{\odot}$/yr for 97\% 
of them and to be $10^{-4}$ $\text{M}_{\odot}$/yr for the other 3\%. \paragraph{}

To get the distribution of ejecta masses for core collapse SNRs we need the distribution for the initial masses of the massive stars. We use the high-mass regime of the initial mass function, that is the distribution function of how massive stars reach the onset of their main-sequence in a given star-formation region~\citep{kroupa_mnras_322_2001}. In the regime of interest it reads: 
\begin{equation}
   dN( M_{\star} ) \propto M_{\star}^{-\zeta}, 
   \label{eq:imf1}
\end{equation}
with $\zeta=2.3$. Hence, the total number of progenitors obeys: 

\begin{equation}
   N = \int_{8}^{M_{\rm max}} dN( M_{\star} ) \propto \int_{}^{}  M_{\star}^{-\zeta} dM_{\star}, 
   \label{eq:imf2}
\end{equation}
with $M_{\rm max}=120\, \rm M_\odot$. By normalising Eq.~\ref{eq:imf1} to unity, one can invert Eq.~\ref{eq:imf2} to have a mass for each progenitor $M_\star(N_i)$. Using this initial mass distribution for high-mass stars, we associate each of our synthetic SNRs to a progenitor star mass.

\paragraph{}
Furthermore, the relation between the zero-age main-sequence masses of high-mass stars 
and their ejecta masses is estimated on the basis of tabulated evolution models from the {\sc geneva} library\footnote{https://www.unige.ch/sciences/astro/evolution/en/database/syclist/}~\citep{2008Ap&SS.316...43E,ekstroem_aa_537_2012}. 
This code calculates the stellar structure in one-dimension including convection and diffusion in the stellar interior, determined with the Schwarzschild criterion for the mixing length theory and the Zahn recipes for diffusion coefficient to the viscosity caused by horizontal turbulence and for the transport of angular momentum. The physics of mass-loss rate is described in the work of~\citet{Georgy_2010PhDT}. 
There are similar libraries of stellar evolution models produced by other evolutionary codes, such as the {\sc boost} database~\citep{szecsi_aa_658_2022}, 
produced by the Bonn stellar evolution code~\citep{brott_aa_530_2011a,brott_aa_530_2011b} or the {\sc mist} library~\citep{choi_apj_823_2016,dotter_apjs_222_2016}, calculated using the {\sc mesa} code~\citep{Paxton+2015,Paxton+2018,Paxton+2019}. Such multi-purpose models are widely-used in many aspects of stellar astrophysics, such as: population synthesis~\citep{Fragos+2023}, hydrodynamical simulations~\citep{meyer_515_mnras_2022,meliani_mnras_515_2022, Meyer_etal_2023MNRAS.521.5354M} and particle acceleration simulations~\citep{das_aa_661_2022}. 

\begin{figure}
        \centering
        \includegraphics[width=0.49\textwidth]{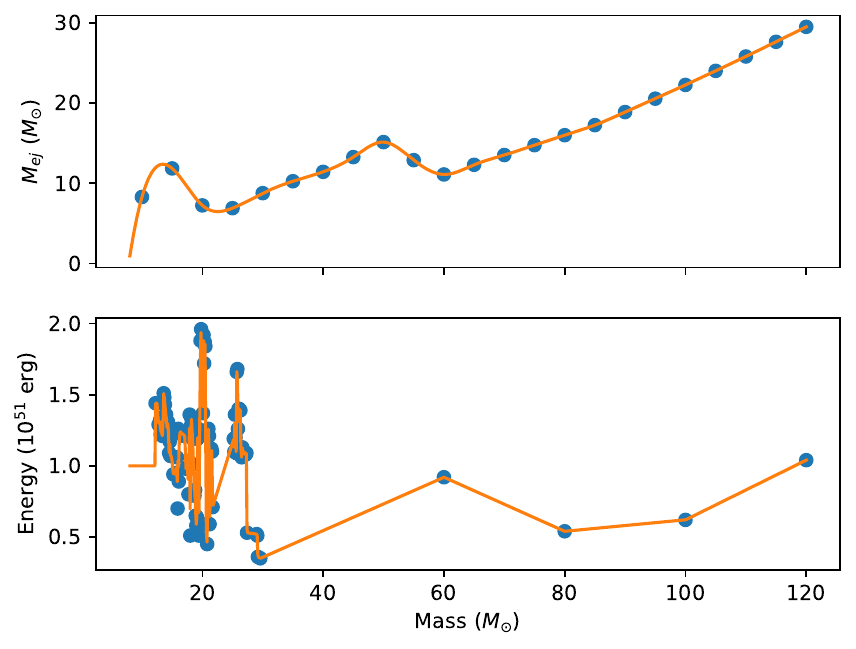}  \\
        \caption{
        Relation between the zero-age main-sequence mass and the ejecta mass 
        of the non-rotating core-collapse supernova progenitors at solar 
        metallicity in the {\sc geneva} library~\citep{ekstroem_aa_537_2012} and the relation between the zero-age main-sequence mass and the explosion 
        energy~\citep{sukhbold_apj_821_2016}. The blue dots show the calculated points while the orange lines show the interpolation between them.
        }
        \label{fig:ZAMS_EJ_masses}  
\end{figure}

\paragraph{}
We use the {\sc syclist} interface to the {\sc geneva} library to generate stellar evolution models for non-rotating  massive stars at solar metallicity ($Z=0.014$), with a zero-age main-sequence mass $M_\star$ spanning from $10$ to $120\, \rm M_{\odot}$. The initial mass range is discretised by a representative number non-rotating stars covering it with a mass increment of $5\, \rm M_{\odot}$. The ejecta mass $M_{\rm ej}$ of these core-collapse supernova progenitors is estimated by subtracting the mass of a neutron star to the pre-supernova mass of the star as calculated by {\sc geneva}. It reads:  
\begin{equation}
   M_{\rm ej} =  M_{\star} - \int_{t_\mathrm{ZAMS}}^{t_\mathrm{SN}} \dot{M}(t)~ dt - M_{\mathrm{NS}},
   \label{eq:co}
\end{equation}
where $t_\mathrm{ZAMS}$ and $t_\mathrm{SN}$ are the zero-age main-sequence and supernova times, respectively, $\dot{M}$ the mass-loss rate of the star at a given time $t_\mathrm{ZAMS} \le t \le t_\mathrm{SN}$ of its evolution and $M_{\mathrm{NS}}=1.4\, \rm M_{\odot}$ is the mass of the neutron star produced after the death of the massive progenitor. The obtained zero-age mass-sequence versus ejecta mass law is shown in Fig.~\ref{fig:ZAMS_EJ_masses} (top) 
and it is implemented into our model via a routine interpolating a table containing the data relative to these $M_\star$-$M_{\rm ej}$ relation.

As the explosion energy $E_{\rm ej}$ is not provided by stellar evolutionary calculations 
but require additional calculations taking them as initial conditions, we use the results 
of~\citet{sukhbold_apj_821_2016} which provide a $M_\star$-$E_{\rm ej}$ law in their 
table 5, referred as model N20K based on neutrino powered explosions and permitting to 
interpolate the progenitor mass to the explosion energy. These data are displayed in Fig.~\ref{fig:ZAMS_EJ_masses} (bottom).

\paragraph{}

 Using these ejecta-mass and explosion energy distributions described above, as well as the uniform age distribution, the SNRs are placed throughout the Milky Way according to one of the following papers: \citet{CAFG_2006}, \citet{Steiman-Cameron_2010}, \citet{Green_2015} or \citet{Reid_2019}. The source distributions will be called CAFG, Steiman-Cameron, Green and Reid from here on and are described in more detail in Sect. \ref{section:CAFG}. The Steiman-Cameron and Green models are both implemented following the work by \citet{Steppa_2020}. In this work we use a Galactocentric coordinate system with the Sun located at $x = 0$, $y = 8.5$~kpc, $z = 0$.

\subsubsection{Spatial distribution of SNRs}

\paragraph{\textbf{The CAFG model:}}

\label{section:CAFG}
The 4-spiral-arm model described in \citet{CAFG_2006} is used to model pulsar positions. We use it by having the birth locations of the supernovae follow the radial distribution described for the pulsars and the arm is chosen at random. The location of the supernovae also follow a $z$ distribution with thermonuclear supernovae following the HI gas density and core collapse supernovae following the HII gas density. The radial distribution of sources uses the following probability density function:
\begin{equation}
    \rho(r) = A \left(\frac{r+R_1}{R_{\odot}+R1} \right)^a \text{exp} \left[-b\left(\frac{r-R_{\odot}}{R_{\odot}+R_1}\right) \right] \, ,
\end{equation}
where $A$ is a normalisation constant, $r$ is the Galactocentric radius, $R_1 = \SI{0.55}{kpc}$, $a = 1.64$ and $b = 0.24$.

\begin{table*}[h]
	\caption{Parameter values for the Steiman-Cameron arms}
	\label{table:Steiman_arms}
	\centering
	\small
	\centering
	\begin{tabular}{|m{1.6cm} *{3}{|m{0.7cm}}| *{2}{m{1.1cm}} *{3}{|m{0.7cm}}|}
	\hline
	Spiral arm & $\beta_i$ & $a_i$ & R (kpc) & \multicolumn{2}{|c|}{$\sigma_r$ (kpc)} & $\sigma_{z,2}$ (kpc) & $\delta$ (deg) & $A_i$ \\
	 & & & & $(r<R)$ & $(r>R)$ & & & \\
	\hline
	Sagittarius-Carina & 0.242 & 0.246 & 2.9 & 0.7 & 3.1 & 0.070 & 15 & 169 \\[8pt]
	Scutum-Crux & 0.279 & 0.608 & 2.9 & 0.7 & 3.1 & 0.070 & 15 & 266 \\[8pt]
	Perseus & 0.249 & 0.449 & 2.9 & 0.7 & 3.1 & 0.070 & 15 & 339 \\[2pt]
	Norma-Cygnus & 0.240 & 0.378 & 2.9 & 0.7 & 3.1 & 0.070 & 15 &	176 \\
	\hline

	\end{tabular}
\end{table*}

\begin{table*}[h!]
	\caption{Parameter values for the arms in the Reid distribution of sources}
	\label{table:Reid_arms}
	\centering
	\small
	\begin{tabular}{|c|c|c|c|c c|}
	\hline
	Spiral arm & $\beta$ Range (deg) & $\beta_{\text{kink}}$ (deg) & $R_{kink}$ (kpc) & \multicolumn{2}{|c|}{$\psi$ (deg)} \\
	 & & & & $(\beta<\beta_{\text{kink}})$ & $(\beta>\beta_{\text{kink}})$ \\
	\hline
	Sagittarius-Carina & $2 \rightarrow 97$ & 24 & 6.04 & 17.1 & 1.09 \\
	Scutum-Centaurus & $0 \rightarrow 104$ & 23 & 4.91 & 14.1 & 12.1\\
	Perseus & $-23 \rightarrow 115$ & 40 & 8.87 & 10.3 & 8.7 \\
	Norma & $5 \rightarrow 54$ & 18 & 4.46 & -1.0 & 19.5 \\
	Local & $-8 \rightarrow 34$ & 9 & 8.26 & 11.4 & 11.4 \\
	3 kpc & $15 \rightarrow 18$ & 15 & 3.52 & -4.2 & -4.2 \\
	Outer & $-16 \rightarrow 71$ & 18 & 12.24 & 3.0 & 9.4 \\
	\hline

	\end{tabular}
\end{table*}

\paragraph{\textbf{The Steiman-Cameron model:}}
 \label{section:Steiman}
A 4-spiral-arm model based on the interstellar medium is described in \citet{Steiman-Cameron_2010}. The model matches the data obtained by the FIRAS instrument of the Cosmic Background Explorer for the [CII] \SI{158}{\micro\metre} and [NII] \SI{205}{\micro\metre} lines of the interstellar medium. The distribution of sources is described by the following probability density equation:
\begin{equation} \label{equation:Steiman}
\begin{aligned}
    \rho (r, \phi, z) = \sum_{i=1}^4 A_i \text{exp} \left( - \frac{1}{\delta^2} \left( \phi - \frac{\text{ln}\left(\frac{r}{a_i}\right)}{\beta_i} \right)^2 \right) \\
    \times \text{ exp} \left( -\frac{|r-R|}{\sigma_r} \right) \text{exp} \left( - \frac{z^2}{2\sigma_{z,2}^2} \right) \, ,
\end{aligned}    
\end{equation}
where $\sigma_r$ is the scale length of the radius, $R$ is the local maximum radius, $\delta$ is the azimuthal scale angle, $\beta_i$ is the pitch angle of the arm, $a_i$ is the orientation of the arm and $\sigma_{z,2}$ is the scale height. The parameters for the arms can be seen in Table \ref{table:Steiman_arms}.

\paragraph{\textbf{The Green model:}}
\label{section:Green}
An azimuthally symmetric model based on SNR data is described in \citet{Green_2015}. The spatial distribution we use is described by the following probability density function:
\begin{equation}
    \rho (r,z) = \left(\frac{r+r_{\rm off}}{R_{\odot}+r_{\rm off}}\right)^{\alpha} exp\left(-\beta \frac{r-R_{\odot}}{R_{\odot}+r_{\rm off}}\right) exp\left(-\frac{|z|}{z_0}\right) \, ,
    \label{equation:green_distribution}
\end{equation}
with $R_{\odot}$ being the distance of the Sun to the Galactic centre, $z_0$ the scale height of the Galactic disc, the shape parameter $\alpha$, and the rate parameter $\beta$. The parameter $r_{\rm off}$ accounts for a non-zero density at $r=0$. Following \citet{Steppa_2020} we set $z_0 = \SI{0.083}{kpc}$, $r_{\rm off} = \SI{0}{kpc}$, $\alpha = 1.09$ and $\beta = 3.87$.

\paragraph{\textbf{The Reid model:}}
 \label{section:Reid}
A 4-spiral arm model based on data from massive stars with maser parallaxes is described in \citet{Reid_2019}. This model also includes some additional features: the Local arm, a 3~kpc arm around the centre of the Galaxy and an outer arm, which connects with the Norma arm. The local arm is described as an arm segment and does not link with any of the other arms. The 3~kpc arm is associated with the Galactic bar and might not be a true spiral arm. The spiral arms are described by Eq. \ref{equation:Reid_arms} and Table~\ref{table:Reid_arms}.
\begin{equation}
    \text{ln}\left(\frac{R}{R_{\rm kink}}\right) = -\left(\beta - \beta_{\rm kink}\right) \text{tan}\psi \, ,
    \label{equation:Reid_arms}
\end{equation}
where $R$ is the Galactocentric radius at azimuth $\beta$, $R_{\rm kink}$ is the radius of the kink in the arm at azimuth $\beta_{\rm kink}$ and $\psi$ is the pitch angle.

\subsection{The Gas Distribution} \label{section:Matter_distribution}
For the matter distribution we need to know the hydrogen density of the region we place our sources into. In this work we use the matter distribution described by \citet{Shibata_2011}, which includes an empirical model for the matter distribution. They describe the gas density of interstellar hydrogen ($H_I$ and $H_2$) as a function of Galactocentric distance and height above or below the Galactic plane. This empirical model matches closely to the one used in the GALPROP code \citep{GALPROP}.

\section{Confronting our simulated populations to \\ available H.E.S.S. data} \label{section:simulating_populations}

Following the modelling described in Sect.~\ref{section:pop_model}, we can simulate the Galactic population of SNRs. The typical parameters of interest are: 
\begin{itemize}
    \item $K_{\rm ep}$ : the electron-to-proton ratio
    \item $\alpha$ : the spectral index of the accelerated particles
    \item $T_{\rm ST}$ : the duration of the ST phase
    \item $\eta$ : the CR acceleration efficiency 
    \item The spatial distribution of sources
\end{itemize}

We typically compute 100 realisations for a given set of parameters, to ensure stability of our results.  Note that the positions of simulated SNRs change for each of the 100 realisations for a given set of properties but these 100 different position sets are constant when changing the population properties other than the source distribution model. This allows us to closely examine how each of the properties affects the population.

\begin{figure}[h]
    \centering
    \includegraphics[width = 0.49\textwidth]{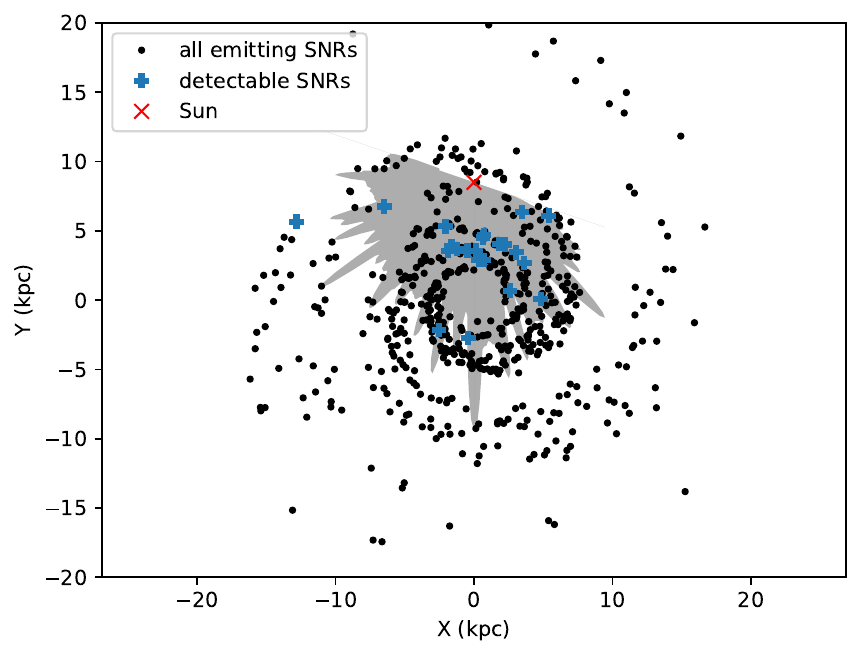}
    \caption{Example of one realisation of a simulation of the population of Galactic SNRs. The gray shaded region is the HGPS detectability range for point like sources with a luminosity of $5\times 10^{33}$ photons $\text{s}^{-1}$ (equivalent to a differential flux at 1 TeV of $\sim 4 \times 10^{-11} \rm TeV^{-1} cm^{-2} s^{-1}$ at a distance of 1 kpc). The blue plusses mark the simulated sources which are detectable and the black dots mark the simulated sources which are not detectable. 3.7\% of the emitting SNRs are detectable in this realisation. The location of the Sun is marked by the red cross.}
    \label{fig:Steiman_distribution}
\end{figure}

\paragraph{}
For each source, using the gamma-ray luminosity,  size and position we are able to determine whether it would have been detected in the HGPS. To account for the selection bias in the H.E.S.S. catalogue we follow the work in \citet{Steppa_2020} and in particular account for the fact that the HGPS sensitivity varies with position and source extent. For a simulated source to be detectable it must exceed the threshold of $5\sigma$ above the background. This condition reads:
\begin{equation}
F_{\rm min} (\theta_{\rm source}) = \begin{cases}
F_{\rm min,0} \sqrt{\frac{\theta_{\rm source}^2 + \theta_{\rm PSF}^2}{\theta_{\rm PSF}^2}}, & \theta_{\rm source} \leq 1^{\circ} \\
\infty, & \theta_{\rm source} > 1^{\circ}
\end{cases} ,
\end{equation}
where $F_{\rm{min,0}}$ is the point source sensitivity, $\theta_{\rm{source}}$ is the source extent (radius) and $\theta_{\rm PSF}$ is the size of the H.E.S.S. point-spread function. The limited field of view of H.E.S.S. combined with the applied technique of deriving background measurements means that sources $\gtrsim 1^\circ$ are not detectable. It is worth noting that the source in the HGPS with the largest extent, Vela Junior, had a dedicated analysis and did not use the analysis method used in the HGPS. Vela Junior has an extent of $1^\circ$, the largest extent of any source analysed using the method in the HGPS is Vela X with an extent of $0.58^\circ$. Another well-known SNR, RX J1713.7-3946, also had a dedicated analysis and has an extent of $0.5^\circ$.
In Fig. \ref{fig:Steiman_distribution} the HGPS detectability range for point like sources with a luminosity of $5 \times 10^{33} \text{photons s}^{-1}$ is displayed, demonstrating the large variations in sensitivity as a function of Galactic longitude. This variation in sensitivity shows why it is not sufficient to have a cut-off in the integrated flux to determine if a source is detectable, and thus why we need to include the sensitivity of H.E.S.S. when testing each simulated source.

\paragraph{}
Because many of the HGPS sources are unidentified and at least some of those should be SNRs, we create an upper and lower limit for the HGPS sources we compare to. The lower limit is the firmly identified SNRs in the HGPS (8) and the upper limit includes all the unidentified sources (47) and the sources identified as composite SNRs (8), this gives us an upper limit of 63 sources to compare to. Moreover, only 12 of the unidentified sources have been found with a position compatible with a known SNR: this allows us to consider a more stringent upper limit of 28 sources. Of the 8 firmly detected SNRs, which have been modelled individually  \citep{HESS_J0852-463, RCW86, HESS_J1534-571, RX_J1713, SNR_G349.7+00.2, HESS_J1731-347, SNR_W28, SNR_W49B}, 4 of them are clearly found with $E_{\rm max} > 10$ TeV, 2 of them with $E_{\rm max} \sim 10$ TeV and 2 of them with  $E_{\rm max} < 10$ TeV. We will use this as an additional constraint on the populations.

\begin{figure*}[h]
\includegraphics[width = \textwidth, page = 2]{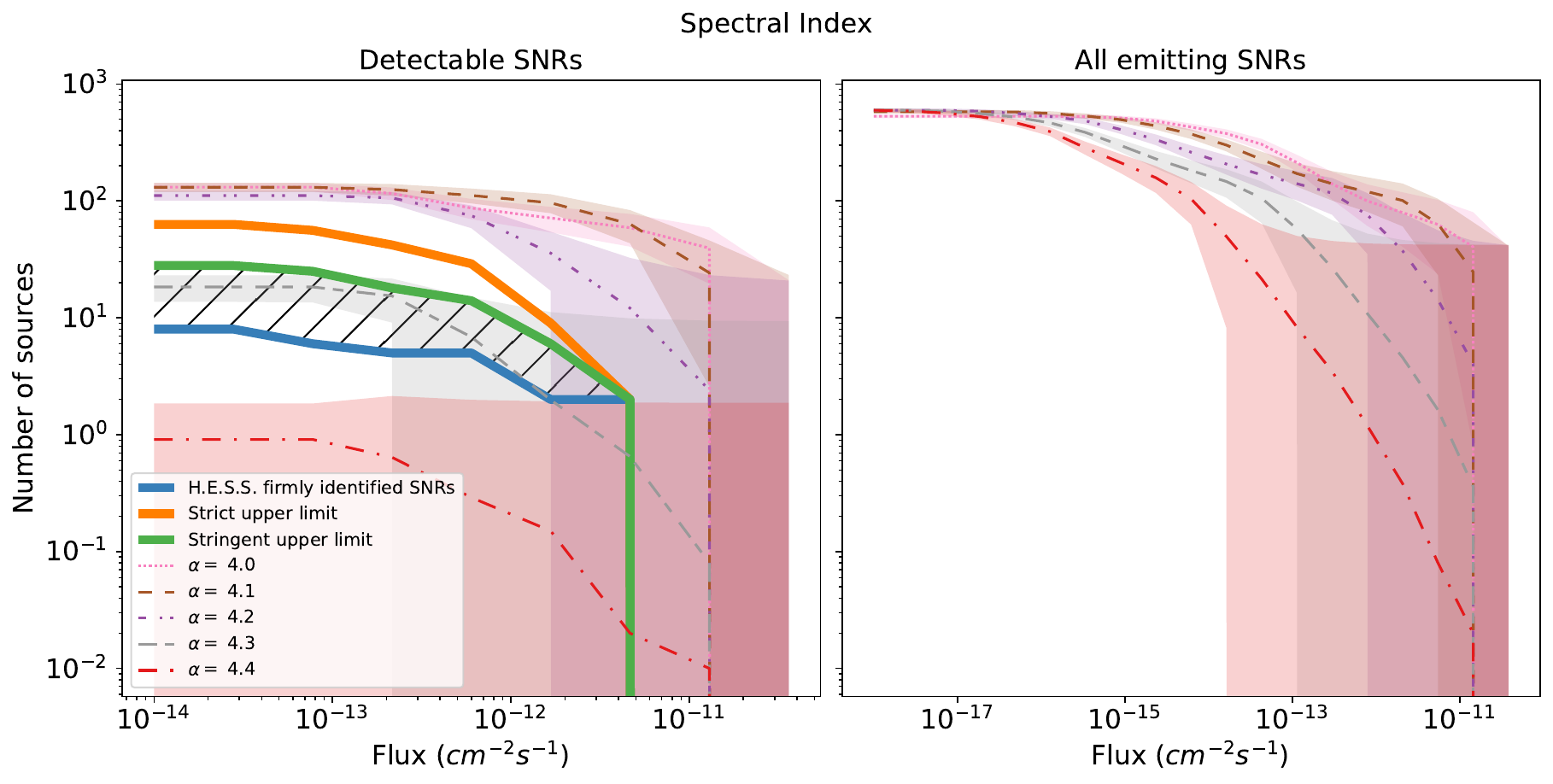}
\caption{
\label{fig:flux_alpha}
$\log N - \log S$ distribution of detectable SNRs (left) and all emitting SNRs (right) based 100 Monte Carlo realisations following a Steiman-Cameron source distribution, $E_{\rm max}$ estimation considering Bell instabilities, $K_{\rm ep}$ = $10^{-3.0}$, and $\eta = 7\%$. The spectral index of the accelerated particles $\alpha$ varies from 4 to 4.4. Shaded areas indicate +/- one standard deviation. Thick solid lines correspond to the HGPS limits (blue: identified SNRs, green: identified and unidentified sources associated to SNRs, orange: all unidentified sources). The hatched region shows the ideal region for the simulated SNRs to be in: between the identified SNRs and the sources associated to SNRs.}
\end{figure*}

\section{Results} \label{section:results}
 We compare our synthetic populations to the HGPS sample. 
For illustrative purposes, we define a reference set of parameters, considering the Steiman-Cameron source distribution, the Bell method to estimate $p_{\rm max}$, $\alpha= 4.3$,  $K_{\rm ep}=10^{-3}$, $\eta = $7\%, and $T_{\rm ST}=$ 20 kyr. It is important to note that this particular set of properties is not the only set that can describe the data but it is a useful as a starting point for making comparisons.

The effect of varying $\alpha$ from  4.0 to 4.4  is shown in Fig.~\ref{fig:flux_alpha}. In this case only one of the sets of parameters (with a spectral index of 4.3) is found in agreement with the data. With a spectral index of 4.3 almost all (96\%) the populations fit between the 8 firmly detected SNRs (blue line) and the upper limit (28) including only sources associated with SNRs (green line) and all the populations fit within the extreme upper limit including all the unidentified sources (orange line). For $\alpha > 4.3$ there are too few detectable SNRs while  $\alpha <4.3$ overproduces the H.E.S.S. results. The log N-log S distribution of fluxes for the simulated SNRs has a very similar shape to that of the H.E.S.S. SNRs. Harder spectra ($\alpha$ close to 4) increase the number of detectable SNRs, which is expected since harder spectra correspond to an increase in the flux in the TeV range, thus increasing the horizon of detectability. The total number of SNRs remains the same for all the  values of the spectral index (right panel Fig.~\ref{fig:flux_alpha}).   

\paragraph{}

\begin{figure*}[h]
\includegraphics[width = \textwidth, page = 4]{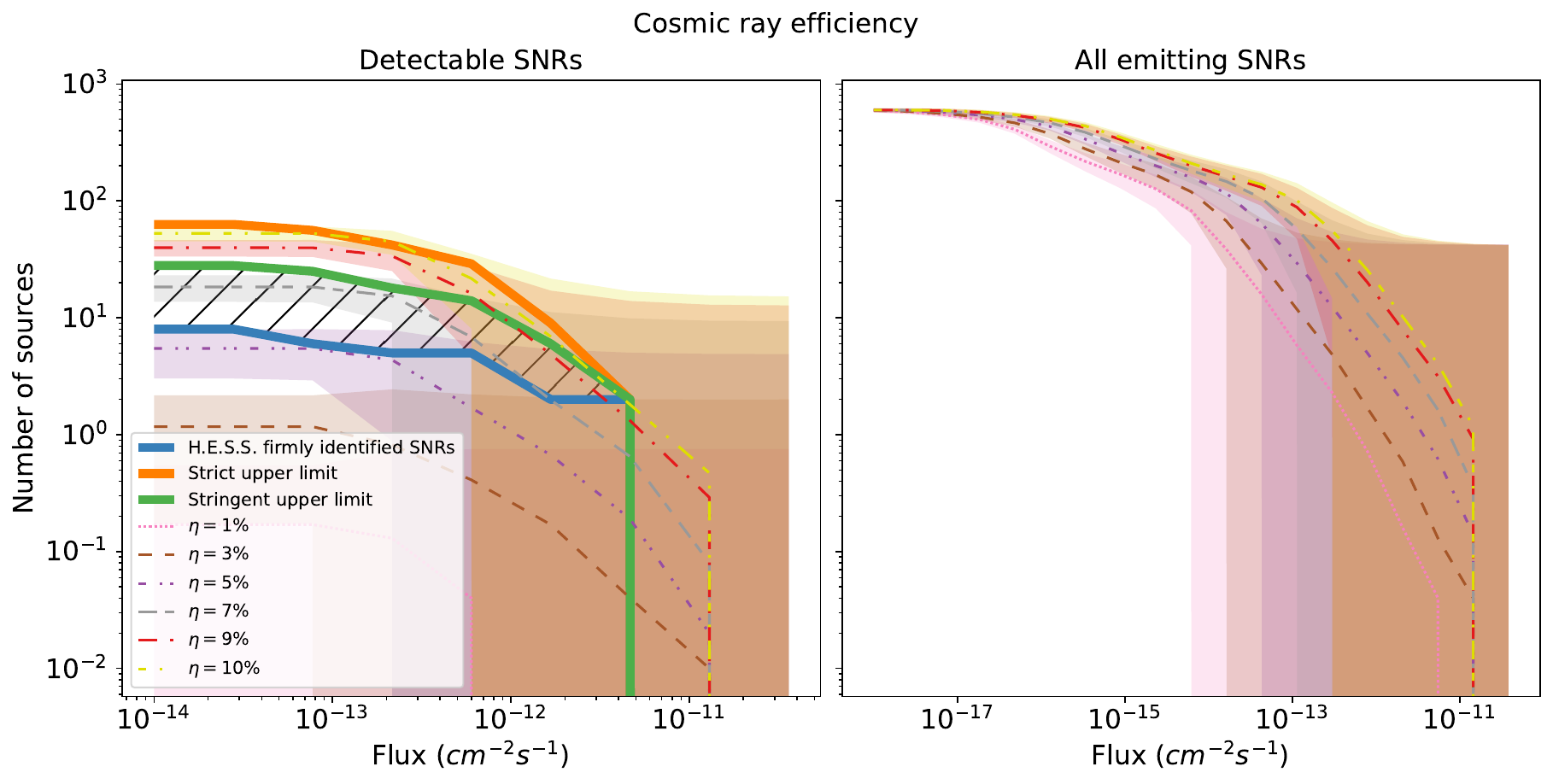}
\caption{
\label{fig:flux_eta}
As in Fig.~\ref{fig:flux_alpha} fixing $\alpha=4.3$ and varying $\eta$ in the range 1\%- 10\%. }
\end{figure*}

Changing the efficiency, $\eta$, from 1\% to 10\% directly impacts the number of detectable SNRs, as shown in Fig.~\ref{fig:flux_eta}. With our benchmark case, we find that $\eta < 5\%$  and $\eta > 9\%$ seem to under or over produce the H.E.S.S. data respectively.

\paragraph{}

\begin{figure*}[h]
\includegraphics[width = \textwidth, page = 3]{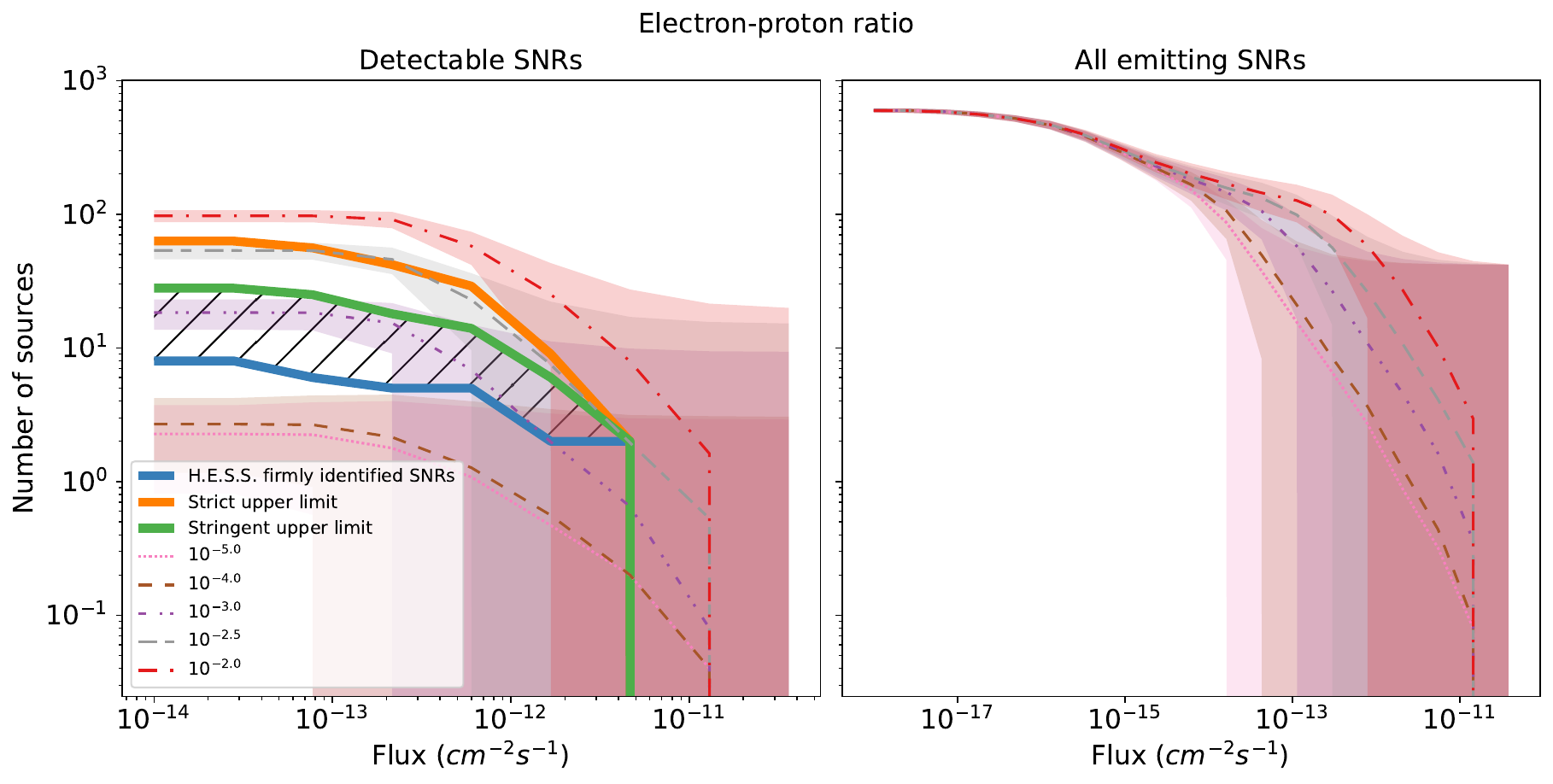}
\caption{As in Fig.~\ref{fig:flux_alpha} fixing $\alpha=4.3$ and varying the electron-proton ratio in the range $10^{-2}$ - $10^{-5}$.}
\label{fig:flux_Kep}
\end{figure*}

Exploring values of $K_{\rm ep}$ in the range $10^{-2}$ to $10^{-5}$, we illustrate in Fig.~\ref{fig:flux_Kep} that a value of $10^{-3.5}$ leads to results compatible within the firmly detected H.E.S.S. SNRs and our stringent upper limit. With the initial benchmark set of parameters we considered, a value of $K_{\rm ep} = 10^{-2.5}$ leads to a simulated population of SNRs close to the upper limit of all the unknown H.E.S.S. sources, thus a possible but extreme and quite unlikely case. 

\begin{figure*}[h]
     \subfigure{\includegraphics[width=0.5\textwidth]{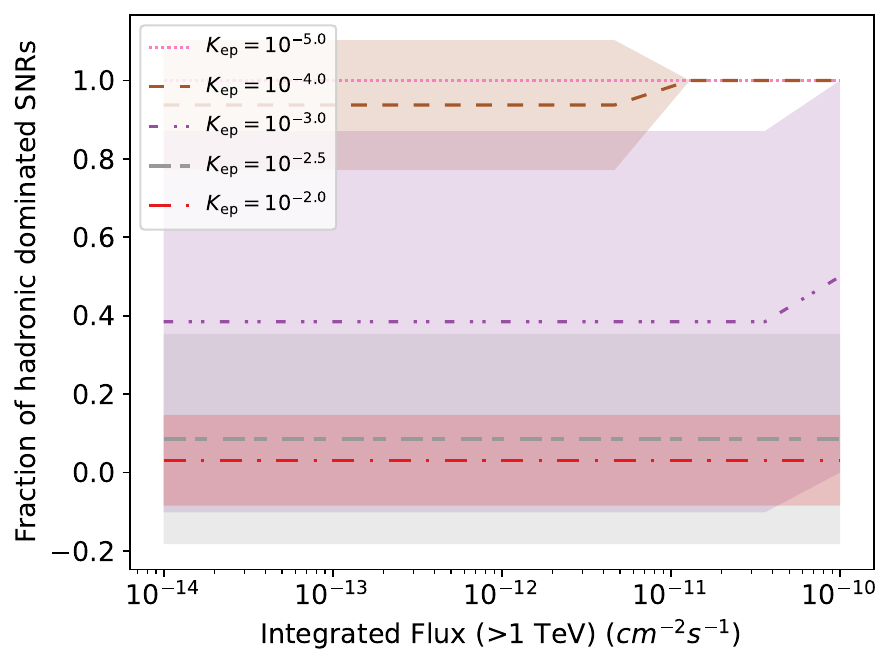}}
     \subfigure{\includegraphics[width=0.5\textwidth]{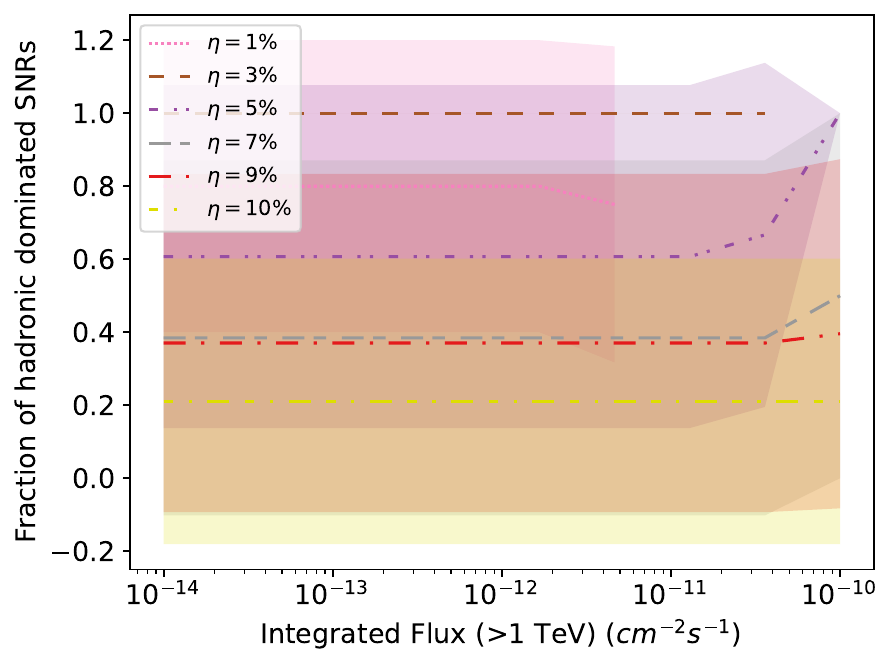}}
     \subfigure{\includegraphics[width=0.5\textwidth]{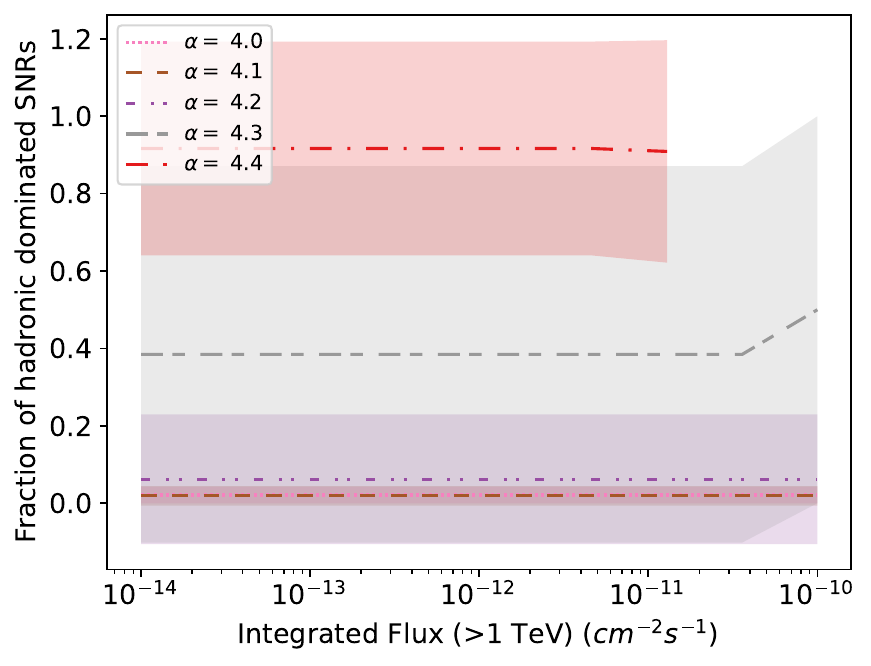}}
    \caption{The mean ratio of hadronic dominated detectable SNRs vs. integrated flux as in Fig.~\ref{fig:flux_alpha}. Top left panel illustrates the importance of $K_{\rm ep}$ (fixing $\alpha = 4.3$ and $\eta=7\%$), top right panel illustrates the effect of varying $\eta$ (fixing $K_{\rm ep} = 10^{-3}$ and $\alpha=4.3$), and bottom panel $\alpha$ (fixing $\eta = 7\%$ and $K_{\rm ep}=10^{-3}$)
    The rest of the properties are as in Fig~\ref{fig:flux_alpha}.}
    \label{fig:had_ratio}
\end{figure*}

\paragraph{}
Fig.~\ref{fig:had_ratio} shows the mean hadronic ratio (ratio of the number of SNRs with gamma-ray luminosity dominated by the hadronic component to the total number of SNRs) against the integrated flux (>1 TeV) of the detectable SNRs using the same cumulative binning as in Figs. \ref{fig:flux_alpha}-\ref{fig:flux_Kep}, and its strong dependence on the parameters $K_{\rm ep}$, $\eta$, and $\alpha$. The dependence on $K_{\rm ep}$ is easily understood since this parameter directly affects the normalisation of the electron spectrum, and thus an increase in $K_{\rm ep}$ leads to a decrease of the hadronic ratio (increase in the number of leptonic dominated SNRs), which can be seen in Fig.~\ref{fig:had_ratio} (top left). At higher integrated fluxes we have fewer detectable SNRs, for $K_{\rm ep} = 10^{-4}$ and $K_{\rm ep} = 10^{-5}$ we have on average <1 of the brightest SNRs as seen if Fig. \ref{fig:flux_Kep}, this brightest detectable SNR is always dominated by hadronic emission and this is why there is no statistical error at the edge of Fig. \ref{fig:had_ratio}. An increase in $\eta$  and/or decrease in $\alpha$ also leads to a decrease of the hadronic ratio (increase of the number of leptonic dominated SNRs). This can be understood as the result of several combined effects: by decreasing $\alpha$ and/or increasing $\eta$, in a given horizon, we allow for a part of the SNR population (older SNRs, or located in less dense parts) to become detectable, and thus unveil a part of the SNR population that is typically more leptonic dominated (\emph{cf.} Table~\ref{tab:information_table}), on top of the already detected hadronic SNRs, thus leading to a decrease of the hadronic ratio.

\paragraph{} 

\begin{figure*}[h]
\includegraphics[width = \textwidth, page = 5]{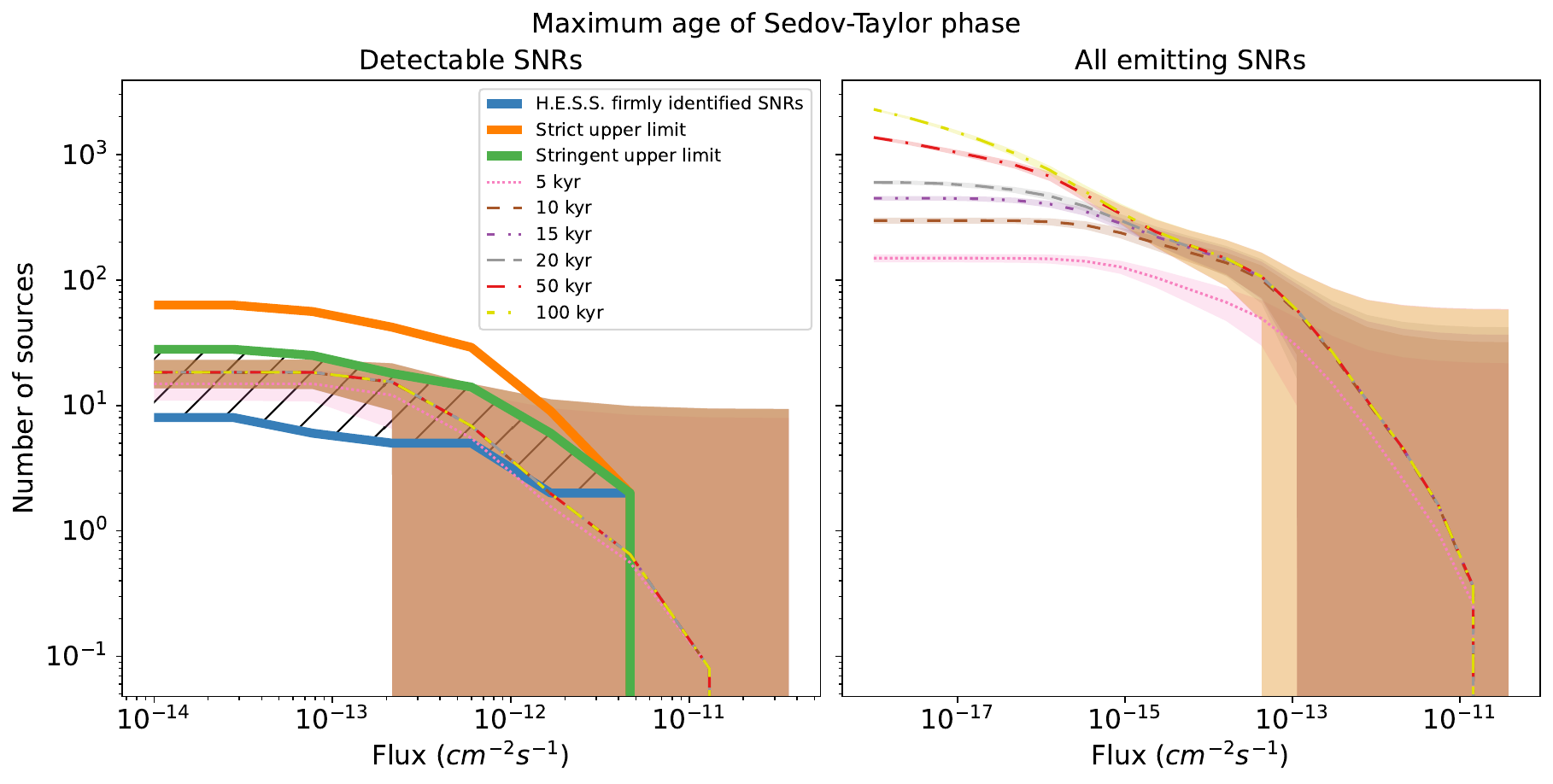}
\caption{
\label{fig:flux_ST}
As in Fig.~\ref{fig:flux_alpha} fixing $\alpha=4.3$ and varying $T_{\rm ST}$ from 5 kyr to 100 kyr.}
\end{figure*}

We additionally investigate the influence of the duration of the active acceleration phase, treating the duration of the ST phase $T_{\rm ST}$ as a free parameter that we are artificially changing. 
Changes within a factor of a few typically do not affect substantially our results, as shown in Fig.~\ref{fig:flux_ST}. This is somehow expected since mostly young $\lesssim 5-10$ kyr old SNRs account for the brightest TeV SNRs. 
\paragraph{}
Diffusive shock reacceleration, as discussed in Sect.~\ref{sec:reac}, is found to be especially relevant at old ($\gtrsim 10$ kyr) SNR shocks~\citep{bell1978,blasi2004,blasi2017} and for small $\alpha$, not impacting substantially our results in the TeV range.

\begin{figure*}[h]
\includegraphics[width = \textwidth, page = 1]{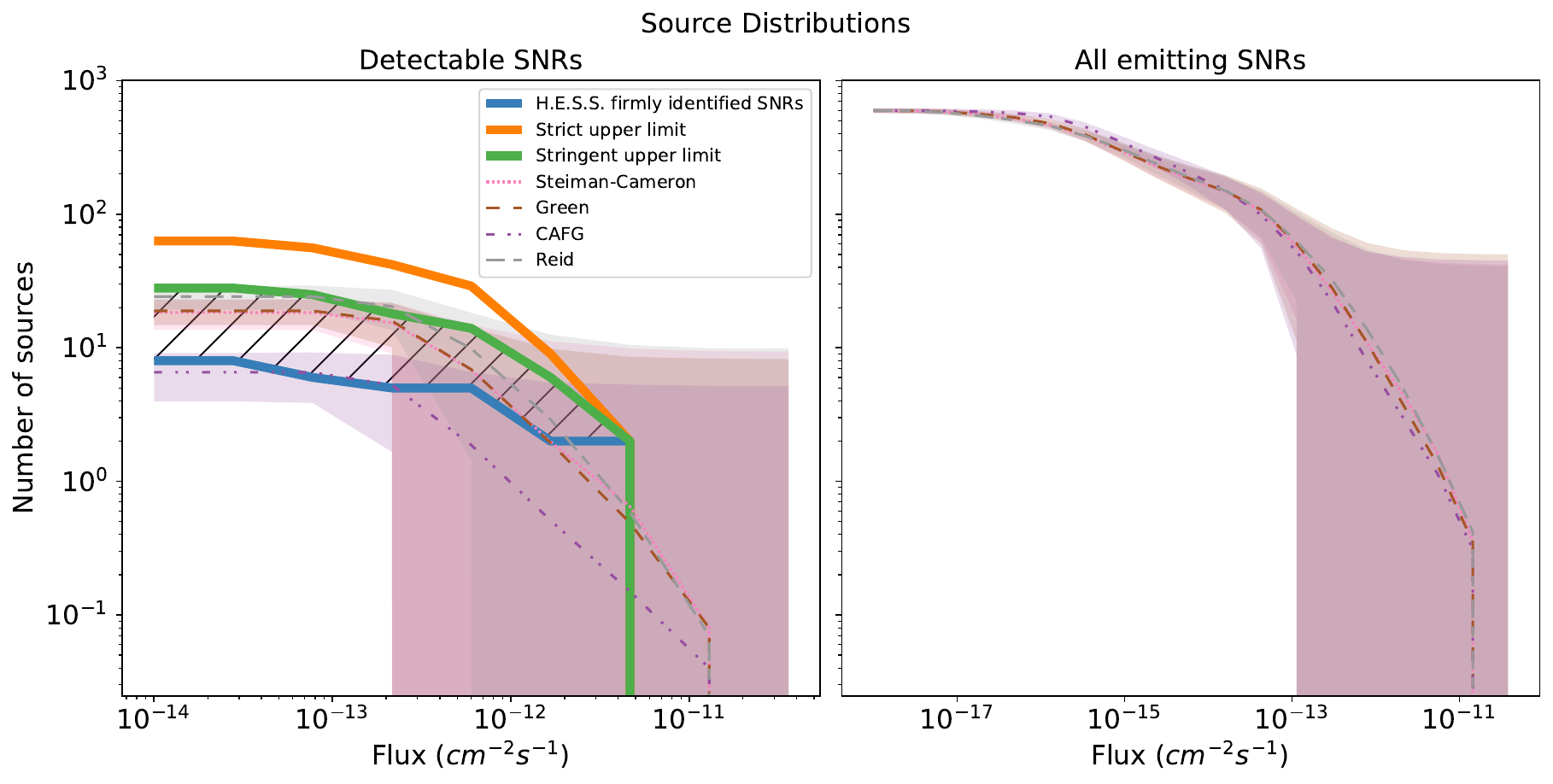}
\caption{
\label{fig:flux_distribution}
As in Fig.~\ref{fig:flux_alpha} fixing $\alpha=4.3$ and varying the source distribution models: Stemian-Cameron, CAFG, Green, and Reid. }
\end{figure*}

\paragraph{}
The different spatial distributions of SNRs are compared in Fig.~\ref{fig:flux_distribution}. The importance of the source distribution is highlighted in the fact that the differences in the number of the detectable subsample of sources between the spatial distributions (left panel) is larger than when considering all emitting sources (right panel). At the same time, this shows that a thorough discussion on the spatial model requires taking into account the refined description of the sensitivity of the instrument, including its spatial dependence as done in this work. Interestingly, the Steiman-Cameron and the Green distributions produce very similar detectable simulated SNRs despite having very different source distributions. 
All spatial distributions considered can account for the HGPS data when adjusting the other parameters of the SNR population.

\paragraph{\textbf{Systematic exploration of the parameter space:}}
In addition to varying each parameter at a time for the benchmark case discussed above, we perform a systematic exploration of the parameter space with $ 4.0 \leq \alpha \leq 4.4$, $10^{-5} \leq K_{\rm ep} \leq 10^{-2}$ and $ 0.01 \leq \eta \leq 0.1$, considering $E_{\rm max}$ from the Bell and Hillas estimates. For these simulated properties the maximum length of the ST phase was set to 20 kyr, and the Steiman-Cameron source distribution were used.

Exploring the parameter space, we constrain the results of our simulations by the HGPS catalogue, especially investigating which parameters can lead to realisations compatible with the detected Galactic population. 

Moreover, we check that the maximum energy of accelerated particles, $E_{\rm max}$, at the simulated SNRs is compatible with the current $E_{\rm max}$ estimated from the measured gamma--ray differential spectrum at each Galactic SNR. Schematically, out of the 8 firmly detected SNRs, 4 of them are clearly found with $E_{\rm max} > 10$ TeV (HESS J0852-463, HESS J1442-624, HESS J1713.7-3946 and HESS J1731-347), 2 of them with $E_{\rm max} \sim 10$ TeV (HESS J1534-571 and HESS J1718-347). Let us mention that the cases of HESS J1801-233 (W28) \citep{W28_2018} and HESS J1911+090 (W49B) \citep{W49_2018} are rather tricky as these two sources are thought to be SNRs interacting with molecular clouds, so it is not easy to ascribe the gamma-ray emission to the SNR shock, or to clouds, or to a combination of the two. In addition, the question of whether there is an actual exponential suppression in the gamma-ray flux is yet not clearly answered. It remains that the gamma-ray luminosity is steeply decreasing in the $\sim 1-10$ TeV range, and we consider in this work that $E_{\rm max} <10$ TeV.

\begin{figure}[h]
 \centering
    \includegraphics[width=0.49\textwidth]{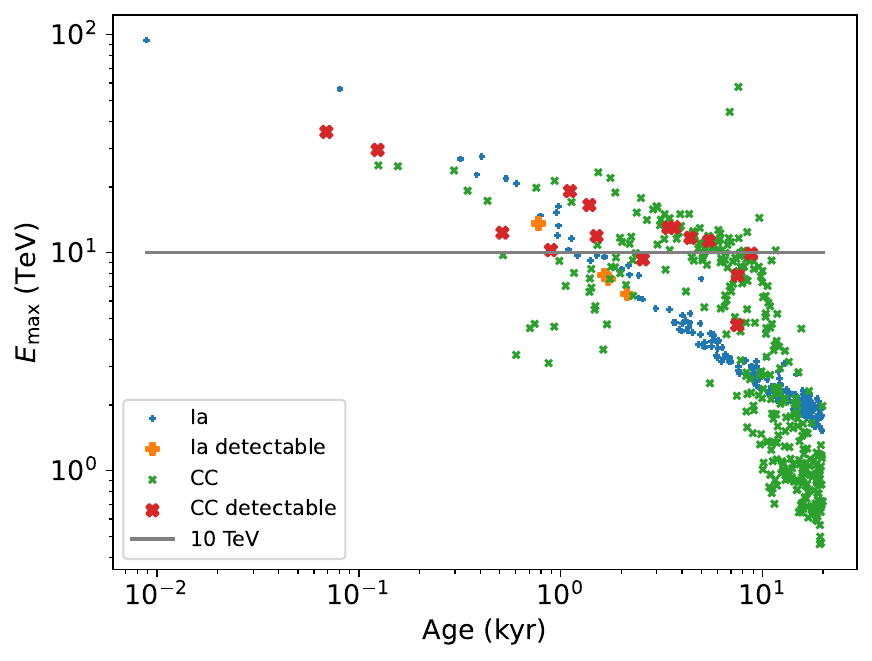}
    \caption{
   Maximum energy of accelerated protons vs. age of each simulated SNR, for a given Monte-Carlo realisation. Parameters adopted are: $\alpha = 4.2$, $K_{\rm ep}=10^{-5}$, $\eta = 9\%$, and $E_{\rm max}$ following the Bell prescription. The horizontal line at 10 TeV illustrates the criterion required to account for the HGPS data.}
           \label{fig:Emax}
\end{figure}

\begin{figure}[h]
 \centering
    \includegraphics[width=0.49\textwidth]{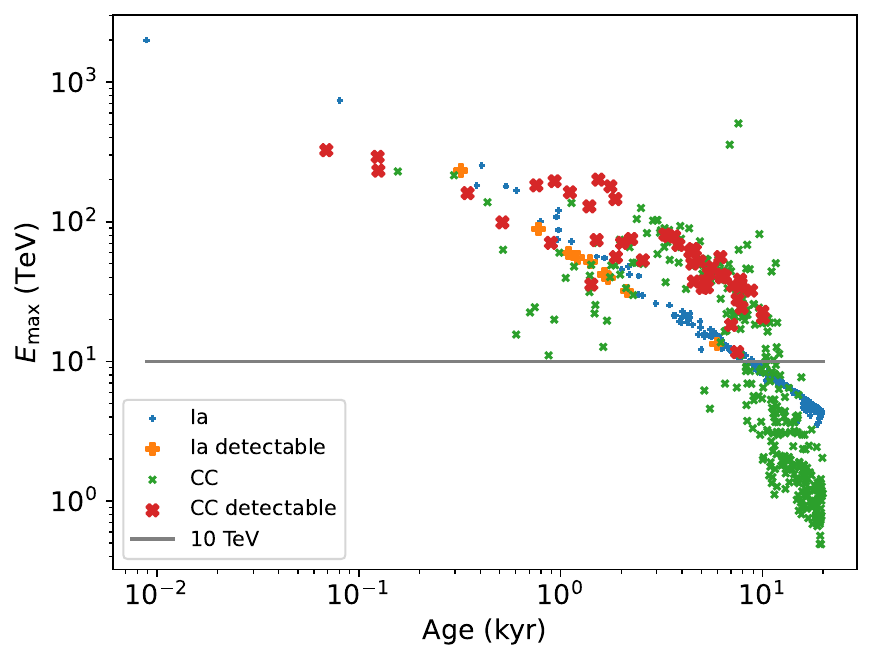}
    \caption{
   As in Fig.~\ref{fig:Emax} but with $E_{\rm max}$ following the Hillas prescription.}
           \label{fig:Emax_Hillas}
\end{figure}

For all realisations of our simulations, we check  the number of realisations for which
1) results are found between the firmly detected H.E.S.S. SNRs (8) and all the detected H.E.S.S. sources associated with SNRs (28) and 2)  $E_{\rm max}$ of the simulated SNRs that are detectable in our sample is compatible with the number of SNRs with $E_{\rm max}> 10$ TeV in the HGPS ($\geq 4$). This is illustrated in Fig.~\ref{fig:Emax}, where, assuming an $E_{\rm max}$ following Bell, one realisation of a simulated population is shown in the parameter space of the SNR age and $E_{\rm max}$. The population parameters are chosen such that they lead to the best agreement with HGPS data ($\alpha = 4.2$, $K_{\rm ep}=10^{-5}$, $\eta = 9\%$, with $\sim 97\%$ of the realisations compatible with data). Interestingly, in this situation, the number of SNR PeVatrons (accelerating particles up to the PeV range) is close to 0, compatible with the idea that PeV CRs are not accelerated at typical SNRs~\citep{Cristofari_2020}. 
The prescription of Hillas naturally leads to possible higher values of $E_{\rm max}$, but does not substantially increase the chances of detecting a SNR PeVatron, this can be seen in Fig.~\ref{fig:Emax_Hillas}.

\begin{figure*}[h]
\includegraphics[width = \textwidth]{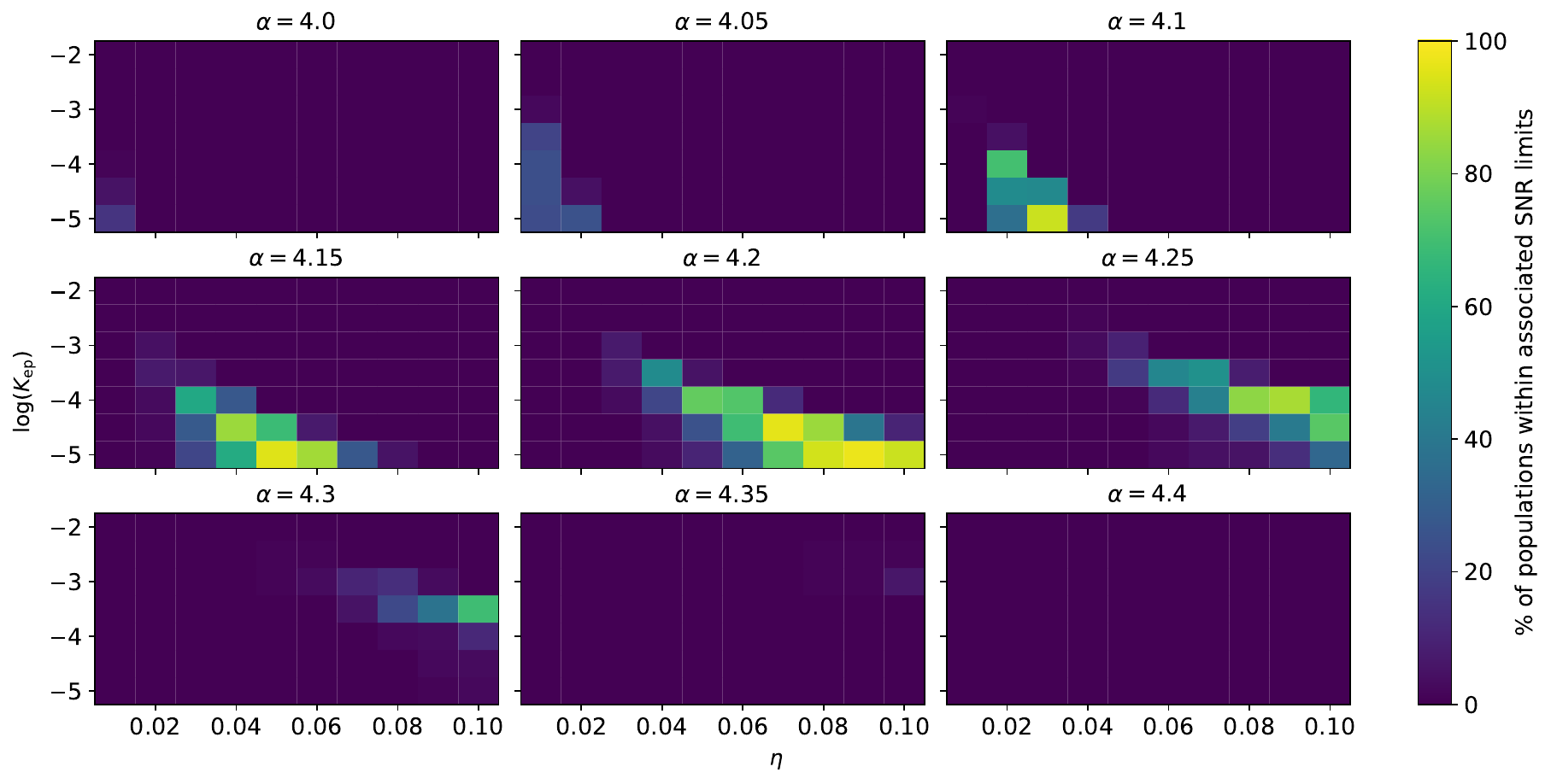}
\caption{
\label{fig:alpha_subplots_Bell}
2D histogram plots showing the percentage of populations that have at least as many detectable simulated SNRs as firmly detected H.E.S.S. SNRs and do not exceed the number of H.E.S.S. sources associated to SNRs, and have at least as many SNRs with $E_{\rm max} \geq 10$ TeV as HGPS SNRs. The source distribution is Steiman-Cameron, and $E_{\rm max}$ following the Bell description.}
\end{figure*}

\begin{figure*}[h]
\includegraphics[width = \textwidth]{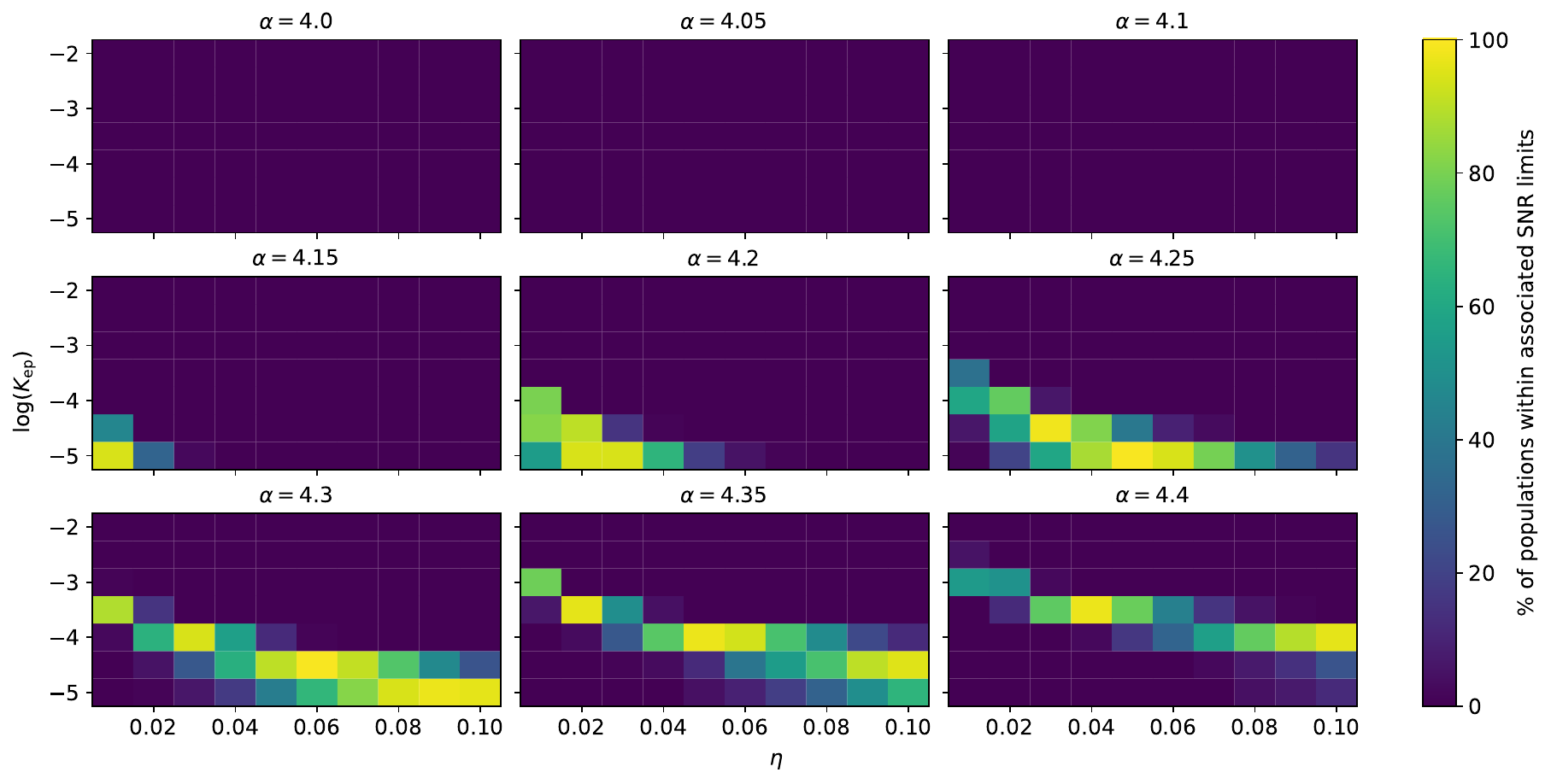}
\caption{
\label{fig:alpha_subplots_Hillas}
As in Fig.~\ref{fig:alpha_subplots_Bell} but with $E_{\rm max}$ estimated with the Hillas criterion.}
\end{figure*}

 The results are illustrated in Fig.~\ref{fig:alpha_subplots_Bell} and Fig.~\ref{fig:alpha_subplots_Hillas}. In Fig.~\ref{fig:alpha_subplots_Bell} the results are displayed assuming $E_{\rm max}$ follows the Bell description, and in Fig.~\ref{fig:alpha_subplots_Hillas} the results with $E_{\rm max}$ estimated with the Hillas criterion.
 
It follows from Fig. ~\ref{fig:alpha_subplots_Bell} that one can rule out the following parameters: $\alpha \gtrsim 4.35$ and $K_{\rm ep} > 10^{-3}$. One can only find a set of parameters for which > 80\% of the realisations are within the associated SNR limits if $4.1 \lesssim \alpha < 4.3$, $K_{\rm ep} < 10^{-3.5}$, and $\eta > 0.02$. 
However, Fig. \ref{fig:alpha_subplots_Hillas} illustrates  that for $E_{\rm max}$ estimated with the Hillas criterion, if $\alpha < 4.15$, or if $ K_{\rm ep}> 10^{-3}$, no region of the parameter space can account for the HGPS data. The lower limit placed on $\alpha$ is a direct result of including the emission from reaccelerated particles into our SNR model. This emission is larger at smaller $\alpha$ and causes the populations with those parameters to have more detectable SNRs than the HGPS. 
On both plots, the correlations between parameters can be seen: an increase in $\eta$ is compensated by either a decrease in $K_{\rm ep}$ or larger values of $\alpha$. 
This intertwining makes it arduous to identify a preferred region of the parameter space.

The systematic exploration of the parameter space allows us to identify the set of parameters that lead to a maximum number of realisations in agreement with the HGPS data (see e.g. Fig.~\ref{fig:alpha_subplots_Bell}, Fig.~\ref{fig:alpha_subplots_Hillas}), but also to examine the properties of these synthetic populations. Assuming an $E_{\rm max}$ following Bell, the typical characteristics of the simulated population for which a maximum number of Monte Carlo realisations are in agreement with the HGPS are shown in Table~\ref{tab:information_table}.

A few remarks are in order. The set of parameters leading to the highest fraction of realisations in agreement with the HGPS is $\alpha = 4.2$, $K_{\rm ep}=10^{-5}$, $\eta=0.09$. For all populations leading to 90\% of compatibility with the HGPS, we found that  $4.2 \gtrsim \alpha \gtrsim 4.1$ and $10^{-4.5} \gtrsim K_{\rm ep} \gtrsim 10^{-5}$ for a CR efficiency $0.1 \gtrsim \eta \gtrsim 0.03$. Interestingly, the electron-to-proton ratio $K_{\rm ep}$ is found to be on the lower side of typically inferred values, and at the same time, typically about $40 \% - 60 \%$ of the gamma-ray emission of the simulated SNRs detectable in the survey is dominated by leptonic interactions (see hadronic ratio). This is somehow counter intuitive, as usually the idea that the VHE gamma--ray emission of SNRs is leptonic implies that the acceleration of leptons is efficient. In our case, low values of $K_{\rm ep}$ are compensated by sufficiently hard spectra and efficiencies, to allow for leptonic dominated gamma rays to be detectable. The leptonic dominated SNRs are typically older than the hadronic dominated SNRs (mean age of leptonic typically twice the mean age of hadronic), consistent with the consensual picture~\citep{pohl1996,reynolds2008,acero2015}, for instance because  of the reduced efficiency of magnetic field amplification at slower shocks, thus reducing losses suffered by electrons. As illustrated in Fig.~\ref{fig:had_ratio}, the efficiency of particle acceleration is indeed found to strongly impact the fraction of leptonic sources. Indeed, the spectra of both accelerated protons and electrons are directly proportional to the acceleration efficiency, but an increase of the efficiency allows for the dimmer leptonic-dominated sources to reach the sensitivity of the HGPS, and thus unveil the leptonic dominated SNRs. With $\alpha =4.3$, and $K_{\rm ep}=10^{-3}$, varying the CR efficiency from 0.01 to 0.1 changes the fraction of hadronic sources from $\approx 1$ to $\approx 0.2$. 
It should be noted that this outcome was achieved assuming that the maximum energy at SNRs is determined by the development of non-resonant streaming instabilities, and ensuring that the count of simulated SNRs detected above approximately 10 TeV aligns with the HGPS data. When considering a maximum energy determined by the Hillas criterion, or relaxing the constraint on the number of SNRs detected above about 10 TeV, it opens up a wider range of possibilities where parameter combinations with an $\alpha$ value around 4.4 or an $K_{\rm ep}$ value around $10^{-2}$ can be identified. Both population ~\citep{Ellison_2007, Cristofari_2013} and theoretical studies~\citep{corso_2023} have not yet lead to stringent constraints for the values of $K_{\rm ep}$ nor the ratio of hadronic dominated sources. \citet{Mandelartz_2015} interpreted 21 out of 24 SNRs in the GeV range as hadronic dominated but the work of \citet{zheng_2019} led to more complex discussions.

\begin{table*}[]
    \centering
    \small
	\begin{tabular}{|lcl|ccccccc|}
	\hline
	\multicolumn{3}{|c|}{Population} & \% of realisations & Hadronic  & Mean No. & Mean Had. & Mean Lep. & Mean Had. & Mean Lep. \\
	\multicolumn{3}{|c|}{Parameters} &  compatible & Ratio & detectable & Age $\mathrm{(kyr)}$ & Age $\mathrm{(kyr)}$ &  Dist. $\mathrm{(kpc)}$ &  Dist. $\mathrm{(kpc)}$ \\
 	& & & with HGPS & & SNRs & & & & \\
	\hline
	$\alpha = 4.2$ & $K_{\rm ep} = 10^{-5.0}$ & $\eta = 0.09$ & 97.0 & 0.62 & 16.84 & 2.15 & 4.86 & 5.65 & 4.88 \\
	$\alpha = 4.2$ & $K_{\rm ep} = 10^{-4.5}$ & $\eta = 0.07$ & 96.0 & 0.43 & 16.14 & 1.94 & 4.36 & 5.64 & 4.9 \\
	$\alpha = 4.15$ & $K_{\rm ep} = 10^{-5.0}$ & $\eta = 0.05$ & 95.0 & 0.51 & 16.41 & 2.06 & 5.21 & 5.62 & 4.79 \\
	$\alpha = 4.2$ & $K_{\rm ep} = 10^{-5.0}$ & $\eta = 0.08$ & 93.0 & 0.66 & 13.6 & 2.0 & 4.88 & 5.63 & 5.06 \\
	$\alpha = 4.1$ & $K_{\rm ep} = 10^{-5.0}$ & $\eta = 0.03$ & 92.0 & 0.37 & 19.56 & 2.05 & 5.7 & 5.61 & 4.63 \\
	$\alpha = 4.2$ & $K_{\rm ep} = 10^{-5.0}$ & $\eta = 0.1$ & 92.0 & 0.6 & 20.64 & 2.32 & 4.92 & 5.66 & 4.76 \\
	$\alpha = 4.25$ & $K_{\rm ep} = 10^{-4.0}$ & $\eta = 0.09$ & 87.0 & 0.27 & 19.34 & 1.11 & 4.34 & 5.71 & 5.06 \\
	$\alpha = 4.15$ & $K_{\rm ep} = 10^{-5.0}$ & $\eta = 0.06$ & 86.0 & 0.48 & 22.83 & 2.31 & 5.2 & 5.65 & 4.63 \\
	$\alpha = 4.15$ & $K_{\rm ep} = 10^{-4.5}$ & $\eta = 0.04$ & 85.0 & 0.37 & 15.26 & 1.87 & 4.65 & 5.66 & 4.83 \\
	$\alpha = 4.2$ & $K_{\rm ep} = 10^{-4.5}$ & $\eta = 0.08$ & 85.0 & 0.38 & 22.67 & 2.03 & 4.6 & 5.64 & 4.88 \\
	$\alpha = 4.25$ & $K_{\rm ep} = 10^{-4.0}$ & $\eta = 0.08$ & 83.0 & 0.32 & 13.97 & 0.96 & 4.13 & 5.73 & 5.03 \\
	$\alpha = 4.2$ & $K_{\rm ep} = 10^{-4.0}$ & $\eta = 0.05$ & 76.0 & 0.29 & 15.11 & 1.04 & 4.19 & 5.74 & 5.03 \\
	$\alpha = 4.2$ & $K_{\rm ep} = 10^{-5.0}$ & $\eta = 0.07$ & 74.0 & 0.68 & 10.41 & 1.82 & 4.89 & 5.63 & 5.22 \\
	$\alpha = 4.25$ & $K_{\rm ep} = 10^{-4.5}$ & $\eta = 0.1$ & 74.0 & 0.58 & 10.87 & 1.72 & 4.14 & 5.66 & 5.08 \\
	$\alpha = 4.2$ & $K_{\rm ep} = 10^{-4.0}$ & $\eta = 0.06$ & 73.0 & 0.21 & 24.56 & 1.19 & 4.54 & 5.71 & 5.01 \\
	$\alpha = 4.1$ & $K_{\rm ep} = 10^{-4.0}$ & $\eta = 0.02$ & 70.0 & 0.2 & 19.06 & 1.41 & 4.74 & 5.78 & 4.88 \\
	$\alpha = 4.2$ & $K_{\rm ep} = 10^{-4.5}$ & $\eta = 0.06$ & 69.0 & 0.49 & 11.26 & 1.65 & 4.3 & 5.69 & 4.99 \\
	$\alpha = 4.3$ & $K_{\rm ep} = 10^{-3.5}$ & $\eta = 0.1$ & 69.0 & 0.19 & 18.19 & 0.78 & 4.05 & 5.82 & 5.14 \\
	$\alpha = 4.15$ & $K_{\rm ep} = 10^{-4.5}$ & $\eta = 0.05$ & 68.0 & 0.31 & 25.19 & 2.17 & 4.76 & 5.66 & 4.81 \\
	$\alpha = 4.25$ & $K_{\rm ep} = 10^{-4.0}$ & $\eta = 0.1$ & 66.0 & 0.24 & 25.46 & 1.2 & 4.56 & 5.68 & 5.02 \\
	$\alpha = 4.15$ & $K_{\rm ep} = 10^{-5.0}$ & $\eta = 0.04$ & 61.0 & 0.57 & 10.67 & 1.7 & 5.24 & 5.67 & 5.1 \\
	$\alpha = 4.15$ & $K_{\rm ep} = 10^{-4.0}$ & $\eta = 0.03$ & 60.0 & 0.26 & 14.63 & 1.13 & 4.28 & 5.78 & 4.98 \\
	$\alpha = 4.25$ & $K_{\rm ep} = 10^{-3.5}$ & $\eta = 0.07$ & 50.0 & 0.14 & 26.48 & 0.78 & 4.35 & 5.82 & 5.12 \\
	\hline
	\end{tabular}
    \caption{Typical characteristics of the detectable SNRs in the simulated populations found to be the most compatible with the HGPS results, as described in Fig.~\ref{fig:alpha_subplots_Bell}. (Col. 3) The hadronic ratio is the ratio of the number of SNRs with gamma-ray luminosity dominated by the hadronic component to the total number of SNRs. (Col. 5-8) Had. and Lep. refer to simulated SNRs dominated by hadronic emission and simulated SNRs dominated by leptonic emission respectively. }
    \label{tab:information_table}
\end{table*}

\section{Conclusion} \label{section:conclusion}
Relying on a Monte Carlo approach, we simulated populations of the Galactic SNRs in the TeV range (built on a physically motivated model of the gamma--ray emission from accelerated protons and electrons at SNR shocks) and confronted the results of the simulated populations to available data from the systematic survey of the Galactic plane performed by H.E.S.S..

We performed a systematic exploration of the parameter space defined by:  $K_{\rm ep}$, the electron-to-proton ratio; $\alpha$, the power-law spectral index of the accelerated particles at the SNR shocks;  $T_{\rm ST}$, the duration of the ST phase; $\eta$, the CR acceleration efficiency; and considering different prescriptions for the description of the maximum energy of accelerated particles $E_{\rm max}$ and different prescriptions for the spatial distribution of sources. 
Our conclusions are: 
\begin{enumerate}
    \item We find that the Galactic population of TeV SNRs can be explained by multiple regions of the parameter space.  Given the current population in the HGPS, only some parts of the parameter space seem to be excluded. For instance, $\alpha \lesssim 4.05$ seem to be disfavoured in all cases. $K_{\rm ep} \gtrsim 10^{-2.5}$ is excluded and $K_{\rm ep} \sim 10^{-3}$ requires $\alpha \gtrsim 4.35$ and $\eta \lesssim 0.02$.

\item Despite the limitations of our approach, the systematic exploration of the parameter space allows us to look for the region of the parameter space producing the most realisations within the limits of HGPS sample. In addition to the requirement on the total number of sources, the additional constraint on the maximum energy, ensuring that $E_{\rm max}$ at SNRs is compatible with values estimated at individual SNRs allows to further reduce the allowed parameter space. One possible solution rendering $\sim 97$\% populations in the limits of the HGPS is: $\alpha = 4.2$, $K_{\rm ep}=10^{-5}$, $\eta = 9\%$, assuming $E_{\rm max}$ from the Bell prescription. In this case, the mean ratio of hadronic-to-leptonic dominated SNRs is 0.62, compatible with the current idea that the TeV emission can be leptonic or hadronic for the different known SNRs. Note that the relatively high efficiency $\eta = 9\%$ allows for a non-negligible fraction of leptonic SNRs, older, or located in low-density regions to be detectable. 

\item The sets of parameters that have $\gtrsim 90 \%$ realisations compatible with the HGPS are obtained with $4.1 \lesssim \alpha \lesssim 4.2$, $ 10^{-5} \lesssim K_{\rm ep} \lesssim 10^{-4.5}$ and $0.03 \lesssim \eta \lesssim 0.1$. Interestingly, this leads to a fraction of $\approx 40 - 60 \%$ of leptonic dominated SNRs in the TeV range, along with a reduced fraction of electrons injected at the SNR shocks. This means that our study naturally highlights situations in which we can account for a significant ($\sim 50\%$) fraction of the sources being dominated by the leptonic emission, with a low value of $K_{\rm ep}$ ($\sim 10^{-5}$). It remains that for most SNRs, the dominant processes at play (leptonic vs. hadronic) are still debated, and thus the overall situation for the Galactic population is also unknown~\citep[see e.g. discussion in ][]{zheng_2019}. In fact, even for what is probably the best-studied SNR of the TeV range RXJ1713-3946, numerous works have still not managed to conclude on the dominant mechanism at play~\citep{ellison_2010,Zirakashvili_2010,Morlino_2012,gabici_2016, ohira_2016,cristofari_2021c,fujita_2022}. In other words, the simulated populations are thus directly compatible with a situation where half of the TeV detected SNRs are leptonic dominated, and at the same time, the fraction of injected electrons at SNR shocks is substantially lower than the $\sim 1/100 (\sim 1/50)$
ratio of electrons to protons measured in the local CR spectrum in the $\sim$ GeV range. This would thus be an additional argument in favour of SNRs not being dominant sources of CR electrons~\citep{gabici2019, lipari2019}, and at the same time, that a substantial fraction of SNRs are detected in the TeV range with a dominant leptonic component. Note that this result was obtained under the assumption that the maximum energy at SNRs is set by the growth of non--resonant streaming instabilities, and ensuring that the number of simulated SNRs detected above $\gtrsim 10$ TeV is compatible with the HGPS data. Considering a maximum energy set by the Hillas criterion, or relaxing the condition on the number of SNRs detected above $\gtrsim 10$ TeV leads to more diverse possibilities where sets of parameters with $\alpha \sim 4.4$ or $K_{\rm ep} \sim 10^{-2}$ can be found.

\item Some parameters do not substantially affect our results. For $T_{\rm ST}$ this is coherent with the idea that most TeV emitting SNRs are young ($\lesssim$ few kyr). The various prescriptions for the spatial distribution of sources do not substantially affect our results. 
\end{enumerate}

Any additional constraint on one of the parameters explored in our study would naturally narrow down the parameter space compatible with the HGPS. Moreover, an increase in the sample of the firm SNR detection (8, our lower limit) would help our approach. This increased number of detections is clearly expected with next-generation instruments such as CTA~ \citep{cristofari2017,scienceCTA,CTA_galactic} or LHAASO~ \citep{LHAASO_SNRs}. \citet{cristofari2017, GPSCTA} discussed that CTA could detect up to several hundreds of SNRs in the systematic exploration of the sky. 

Additionally, joint observations of SNRs with next generation radio observatories~\citep{meerkat,MWA,SKA} will bring data of unprecedented value on the content of non-thermal electrons accelerated at SNR shocks, and help disentangle between the hadronic/leptonic content, thus helping us better understand the details of DSA at SNRs.

 \begin{acknowledgements}  
 This work has been funded by the Deutsche Forschungsgemeinschaft (DFG, German Research Foundation) with the grant 500120112 and supported by the grant PID2021-124581OB-I00 funded by 
MCIN/AEI/10.13039/501100011033 and 2021SGR00426 of the Generalitat de Catalunya. 
This work was also supported by the Spanish program Unidad de Excelencia Mar\' ia 
de Maeztu CEX2020-001058-M. PC acknowledges support from the PSL Starting Grant GALAPAGOS. 
The authors gratefully acknowledge the computing time made
    available to them on the high-performance computer "Lise" at the NHR
    Center NHR@ZIB. This center is jointly supported by the Federal
    Ministry of Education and Research and the state governments
    participating in the NHR (www.nhr-verein.de/unsere-partner).

 \end{acknowledgements}

\bibliography{Population_Galactic_SNRs}
\end{document}